\documentclass[aip,prl,amsmath,amssymb,reprint,superscriptaddress]{revtex4-1} %preprint version
\usepackage{graphicx}% Include figure files
\usepackage{dcolumn}% Align table columns on decimal point
\usepackage{bm}% bold math
\usepackage{epstopdf}

\begin{document}
\title{Temporal and Spatial Turbulent Spectra of MHD Plasma and an Observation of Variance Anisotropy}

\author{D.A. Schaffner}
\affiliation{Swarthmore College, Swarthmore, PA, USA}
\author{M.R. Brown}
\affiliation{Swarthmore College, Swarthmore, PA, USA}
\author{V.S. Lukin}
\affiliation{Space Science Division, Naval Research Laboratory, Washington, DC, USA}

\date{\today}
\begin{abstract}
The nature of MHD turbulence is analyzed through both temporal and spatial magnetic fluctuation spectra. A magnetically turbulent plasma is produced in the MHD wind-tunnel configuration of the Swarthmore Spheromak Experiment (SSX). The power of magnetic fluctuations is projected into directions perpendicular and parallel to a local mean field; the ratio of these quantities shows the presence of variance anisotropy which varies as a function of frequency. Comparison amongst magnetic, velocity, and density spectra are also made, demonstrating that the energy of the turbulence observed is primarily seeded by magnetic fields created during plasma production. Direct spatial spectra are constructed using multi-channel diagnostics and are used to compare to frequency spectra converted to spatial scales using the Taylor Hypothesis. Evidence for the observation of dissipation due to ion inertial length scale physics is also discussed as well as the role laboratory experiment can play in understanding turbulence typically studied in space settings such as the solar wind. Finally, all turbulence results are shown to compare fairly well to a Hall-MHD simulation of the experiment. 
\end{abstract}

\maketitle

\section{Introduction}

As instrumentation capabilities on spacecraft have steadily improved, focus in space plasma turbulence has pushed into ion, sub-ion and electron scale regions in both the solar wind~\cite{alexandrova09,sahraoui09} and the magnetosheath~\cite{sahraoui06,yordanova08}. Newer analysis techniques have also been developed to better tap the details of the rich turbulent environments of the heliosphere and magnetosphere. 

%The space turbulence community has seen a flurry of results in recent years, as diagnostic capabilities on spacecraft have steadily improved, especially in temporal resolution. This has allowed for investigation into ion, sub-ion and electron scale turbulence measurements in many space plasma conditions including, for example, the solar wind~\cite{alexandrova09,sahraoui09} and the magnetosheath~\cite{sahraoui06,yordanova08}. In addition, 

Analysis of timeseries fluctuations in the guise of spectrum analysis has served as the foundation for plasma turbulence research, particularly in the solar wind~\cite{goldstein95,tumarsch95}. Improving instrumentation has led to ever greater precision in measurements of spectral indices for both magnetic and velocity fluctuation spectra~\cite{podesta07}, as well as increasing resolution of scale, allowing for measurements of spectra of sub-ion and electron scale fluctuations~\cite{yordanova08,alexandrova09,sahraoui09,chen13}. However, many outstanding questions regarding the nature of this turbulence, including both injection and dissipation mechanisms, have prompted research beyond typical spectral analysis. Two such avenues have been the exploration of the anisotropic nature of the turbulence and comparison of spectra between different observables (i.e. magnetic versus density fluctuations).

Theoretical treatments of magnetized plasma turbulence have almost universally predicted anisotropy to develop based on directions perpendicular and parallel to a mean field vector~\cite{montgomery81,matthaeus90,goldreich95,zhou04,boldyrev06}; many forms of this anisotropy have been predicted and observed~\cite{dasso05,horbury08,podesta09,perri09,wicks10,he11} (a good overview of the various types and definitions of anisotropy is given in Ref. 20. Since the solar wind is super-Alfvenic, anisotropy studies of solar wind turbulence generally reference the velocity vector of the bulk flow when defining perpendicular versus parallel fluctuations---typically resulting in a wavenumber anisotropy measurement. Instead, this paper focuses on variance anisotropy---the unequal ratio of magnetic fluctuation power in components perpendicular versus parallel to the mean magnetic field~\cite{belcher71,smith06}.

Comparison amongst fluctuation spectra (typically magnetic, flow, and density fluctuations) also provide insight into the nature of the turbulence. For example, relative partition of fluctuation energy between magnetic and kinetic can be useful in making determinations of turbulent properties~\cite{podesta07}, while comparison of the spectral indices of magnetic and density spectra can provide information about the compressibility of a plasma(e.g., the so-called density bulge~\cite{coles89,harmon05}). Comparison of velocity and magnetic spectra at injection scales can also be used to develop hypotheses on the generating and dissipation mechanisms of the turbulence~\cite{roberts10}.

As results from space become more detailed, comparisons to simulation and laboratory experiment become more useful in order to better develop the theory as well as inform future space missions. A long gap in laboratory research has occurred since the earliest observations of magnetic turbulence and anisotropy~\cite{robinson71} as turbulence work in experiments gravitated toward electrostatic turbulence and transport in tokamaks~\cite{liewer85,tynan09}, but more recently, much progress has been made to develop laboratory experiments which can inform space plasma turbulence research~\cite{howes12a} as well as begin to make magnetic turbulence measurements~\cite{ren11}. Experiments conducted in the MHD wind-tunnel configuration of the Swarthmore Spheromak Experiment have shown the ability to produce and analyze MHD turbulence using many of the same methods used in the space plasma community~\cite{deWit13}, and have produced turbulence measurements which are comparable to \textit{in-situ} results~\cite{schaffner14a,schaffner14b}, particularly to magnetosheath spectra~\cite{yordanova08} and solar wind intermittency~\cite{sorrisovalvo99}.

A full spectral analysis of fluctuations in the SSX has been conducted, including magnetic field, density and Mach flow fluctuations. Using a wavelet transform method to decompose the timeseries signal, the magnetic field fluctuation spectra can be broken into portions that are parallel or perpendicular to the local magnetic field vector. Analysis of these portions show that parallel fluctuation power decreases slightly faster than that for perpendicular fluctuations generating a separation in power as a function of increasing frequency. Such change in variance anisotropy has been observed in solar wind turbulence~\cite{kiyani13}. The ratio is shown to grow from nearly isotropic levels in the energy injection scale to a maximum $\perp/\parallel$ ratio of 3 at frequencies of about 1MHz. Beyond this point, the ratio begins to return to isotropic levels. Though the axial flow speeds of the plasma in the wind-tunnel are sub-sonic ($M_{z} < 1$), if a Taylor hypothesis were to be invoked, this peak in the ratio would correspond to the ion inertial scale length. 

Since density and Mach fluctuations timeseries are also recorded, comparisons amongst these spectra are reported. Comparison of Mach number fluctuation spectra (as a proxy for flow) to magnetic spectra in the lowest frequencies suggests that energy injection into the turbulent spectra is primarily magnetic. This result is consistent with the nature of the spheromak generation process. The indices of density and magnetic spectra appear to diverge at a scale consistent with the ion inertial length, where anisotropy is also found to decrease. Both results would be consistent with an increase in compressibility of the plasma due to ion scale physics.

Finally, direct wavenumber spectra are made using a multi-channel probe and are compared to time-domain spectra transformed to spatial scales using the Taylor Hypothesis. This diagnostic highlights an advantage of laboratory experiments since actual spatial measurements are extremely difficult {\it in-situ}, though progress is being made using multiple spacecraft~\cite{sahraoui06,sahraoui10}.  The wavenumber spectra tend to be slightly shallower than their time-domain counterparts. Comparison of the wavenumber spectra with respect to flow gives a hint at an observation of wavenumber anisotropy, but the results are not as conclusive compared to the variance anisotropy analysis.

All the temporal and spatial spectra analyses are compared to simulations generated using the HiFi framework~\cite{schaffner14a}. Spectra are generated from synthetic diagnostic timeseries of magnetic fields, flow, and density. Results compare favorably to the experiment including the observation of variance anisotropy.

The organization of this paper is as follows: First, a brief description of the plasma laboratory is given. Then, an overview of the analysis techniques used to determine perpendicular and parallel fluctuations is provided; details of the method and a second method used to verify its validity are described in the Appendix. The main results of this paper are provided in Sections~\ref{sec:variance} to~\ref{sec:wavenumber}: Section~\ref{sec:variance} presents the variance anisotropy results, Section~\ref{sec:flowdens} compares magnetic fluctuation spectra with velocity and density, and Section~\ref{sec:wavenumber} shows the results of the direct wavenumber spectra analysis. Section~\ref{sec:simulation} presents the comparison of simulation with experimental results. Section~\ref{sec:ionscale} discusses the results in light of their relationship to possible ion inertial length scale dissipation. Concluding remarks are given in Section~\ref{sec:conclusions}. The appendices give details of the variance anisotropy analysis as well as a discussion of wavelet versus FFT transforms and local versus global magnetic fields.

\section{Experiment}\label{sec:experiment}

The turbulence injection process in a laboratory experiment is naturally going to have a different origin than a space physics process; however, it is assumed that processes after the energy injection state (i.e. energy transfer in the inertial range and dissipation) will be similar enough so that exploration of the physics behind them in the laboratory can be beneficial to an overall understanding.

\begin{figure}[!htbp]
\centerline{
\includegraphics[width=8.5cm]{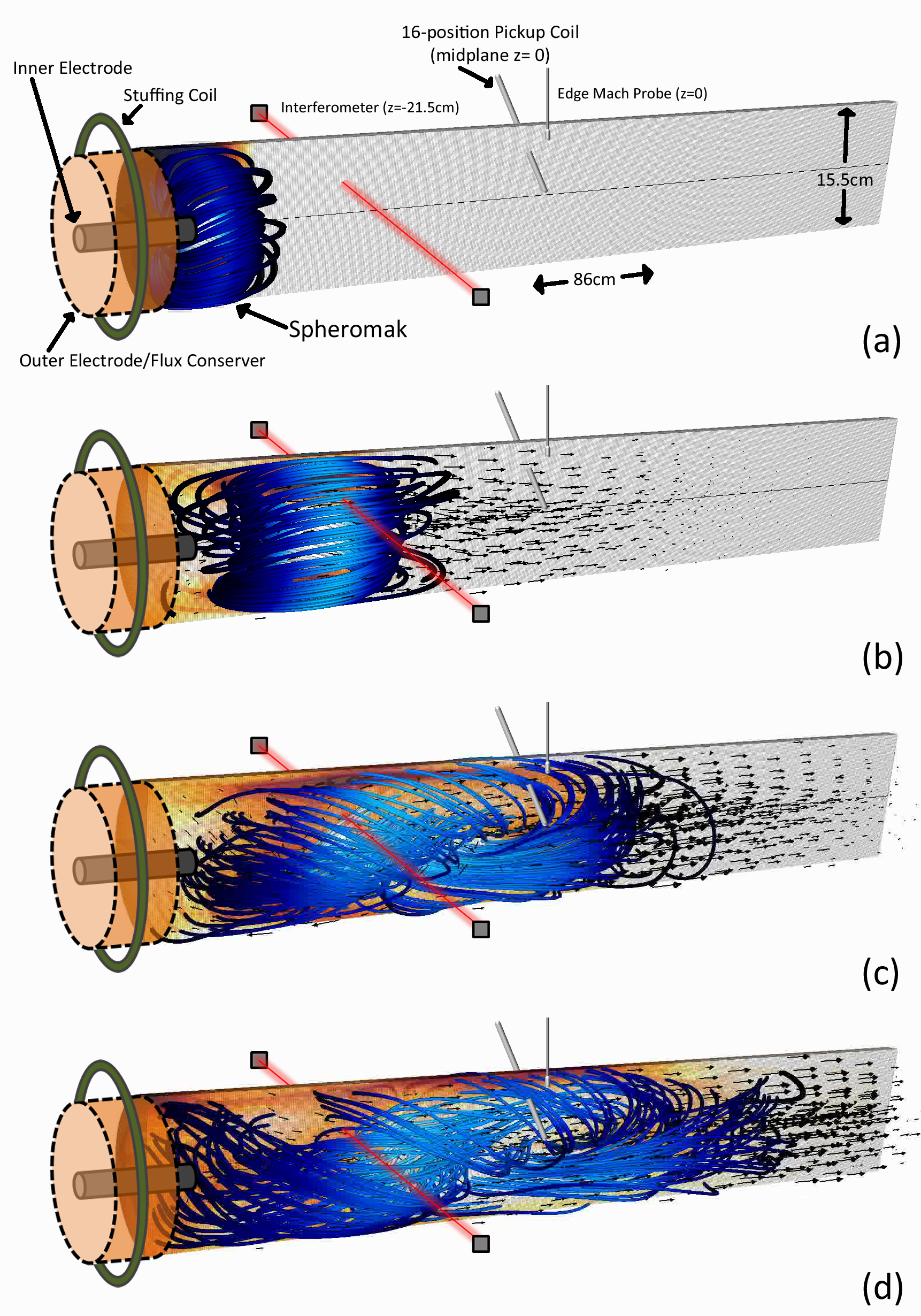}}
\caption{\label{fig:moviestills} A diagram of the MHD wind-tunnel inside the SSX chamber. A plasma gun is located on one end and consists of two electrodes and a stuffing coil. Magnetic pickup coils are located at 16 radial locations at the midplane and a Mach probe measures edge flow at a different angular position at the midplane. The position of the density HeNe interferometer chord is also indicated. Panels (a)-(d) demonstrate the plasma evolution once a spheromak has been formed at the end of the plasma gun(a), and shows the tilt instability(b), twisting under helicity conservation(c) and measurement of turbulent fluctuations(d). The blue lines show simulation generated magnetic field lines, arrows indicate plasma flow magnitude and direction, and orange intensity represents electron density.}
\end{figure}

The energy for the turbulence found in the MHD wind-tunnel in the Swarthmore Spheromak Experiment originates in the plasma production process. As diagrammed in Figure~\ref{fig:moviestills}, a plasma gun configuration sits on one end of a 15.5cm diameter, 86cm long cylindrical copper column which constitutes the MHD wind-tunnel. The gun consists of a tungsten-coated 4cm diameter inner electrode placed concentrically within the copper cylinder which serves as an outer electrode. An axially aligned wire coil surrounds both electrodes and current is supplied to the coil to produce a known amount of magnetic flux---between 0 and 1.5mWb---axially through the inner electrode: this flux is referred to as stuffing flux, $\Phi$. A 1mF capacitor bank, charged to 4kV is discharged across the electrodes; this voltage fully ionizes a small volume of hydrogen gas puffed in just before the discharge. Radial currents through this newly produced plasma push the plasma down the column and into the fringe magnetic fields which tend to resist this push and stuff the progress of the plasma (hence the term, stuffing flux). Given enough current, and thus large enough $J\times B$ force, the plasma distends the stuffing fields until they break off, forming a self-contained magnetic field structure called a spheromak~\cite{barnes86,jarboe93}. This structure is visualized in Figure~\ref{fig:moviestills}(a) using Hall-MHD simulation generated field lines (in blue). Since the spheromak has both polodial and torodial magnetic fields, the relative ratio of field strength between these two directions is quantified by the magnetic helicity, defined as,
\begin{equation}
K_{B} = \int A \cdot B dV
\label{eq:helicity_th}
\end{equation}
where A is the vector potential and dV is the volume element. Previous work on SSX has shown that magnetic helicity of the plasma scales approximately linearly with the amount of stuffing flux applied to the gun~\cite{schaffner14b}.

Figure~\ref{fig:moviestills}(a)-(d) illustrates the experimental procedure. The generalized turbulence cascade begins with this compact magnetic structure (Figure~\ref{fig:moviestills}(a)). Inside the wind-tunnel, the magnetic structure is energetically unstable~\cite{bondeson81,jarboe93}---the structure will begin to tilt over and expand into the remainder of the wind-tunnel (Figure~\ref{fig:moviestills}(b)). Because the column is copper, and thus flux conserving, the magnetic helicity is conserved unlike the magnetic energy~\cite{taylor86}; thus, the structure also begins to twist as it tilts over as seen in Figure~\ref{fig:moviestills}(c). The free energy released in the fall-over materializes as fluctuations in the field, generating the turbulent cascade. The turbulent fluctuations are most prominent in figure~\ref{fig:moviestills}(d). In the actual experiment, the gun typically injects more than a single, self-contained structure so while an initial structure is decaying, more compact field energy is being injected~\cite{barnes86}. This allows for a time frame of stationary fluctuations that is used in the turbulence analysis.

The turbulence data is extracted from magnetic, density and flow measurements during this stationary period. Magnetic fluctuations are recorded using 3mm diameter, single loop pick-up coils located at 16 radial locations along the radius of the midplane of the column as indicated in Figure~\ref{fig:moviestills}(a). Each radial position has three orthogonal loops oriented along the axial, radial and azimuthal directions of the column. A 64MHz, 14-bit DTaq digitizer records $dB/dt = \dot{B}$ timeseries data which is converted into magnetic fluctuation data in frequency space (as discussed in Section~\ref{sec:analysis}). Line-integrated density data is measured with a HeNe interferometer located 21.5cm off of the midplane and flow fluctuations are estimated from a Mach probe located on the edge of the copper column at the midplane (as indicated in Figure~\ref{fig:moviestills}(a)). Spectra of $M_{z}(t)$ are directly reported as a proxy for $V_{z}(t)$, since $\tilde{M}\sim \tilde{V}/C_{s}$ where $C_{s}=(T_{e}/m_{i})^{1/2}$ and is approximately constant with a measured value of $T_{e}=10eV$~\cite{zhang11, schaffner14b}. Bulk flow of the plasma is estimated with time-of-flight measurements between the density signal at z=-21.5cm and the magnetic signal at the midplane, z=0. The plasma is also generated with a set amount of magnetic helicity which is governed by initial conditions of the plasma gun source---namely, amount of flux generated in the gun core (the stuffing flux). The helicity can be scanned~\cite{schaffner14b}, but in this work focus is primarily on two states: a state with non-zero helicity, $K_{B}\neq 0$, generated by 1.0mWb of flux in the gun core and a state with no injected helicity, $K_{B}=0$. Table~\ref{tab:params} indicates typical plasma parameter values for these two states.

%1mWB run
%dens 1.39e15 cm^-3
%bfield 5283 G
%Ti = 23 eV
%bulk flow = 20km/s

%results
%Va = 309km/s
%Beta = 0.07
%Cs = 31km/s
%f_i = 8MHz
%nu_i = 6MHz
%rho_i = 0.09cm
%c/w_pi = 0.61cm
%ion_mfp = 0.16cm
%doppler shifted c/wpi = 3MHz
%doppler shifted rho_i = 20MHz

%0mWb run
%dens 2.84e15 cm^-3
%bfield 747 G
%Ti = 17 eV
%bulk flow = 20km/s

%results
%Va = 30km/s
%Beta = 5.5
%Cs = 31km/s
%f_i = 1.1MHz
%nu_i = 19MHz
%rho_i = 0.56cm
%c/w_pi = 0.43cm
%ion_mfp = 0.05cm
%doppler shifted c/wpi = 4.3MHz
%doppler shifted rho_i = 3.3MHz

\begin{table}
\caption{\label{tab:params}MHD wind tunnel plasma parameters during the equilibrium epoch for the present configuration of SSX for non-zero helicity ($K_{B}\neq 0$) and for zero helicity($K_{B}=0$) states. The table has separate sections for directly measured parameters and for quantities computed from these values.}
\begin{tabular}{c|cc}
\toprule
Parameter&$K_{B}\neq 0$&$K_{B}=0$\\
				 &(1.0mWb)             &(0.0mWb)\\
\hline
$Measured$&&\\
\hline
$\langle |B|\rangle [kG]$&5.283&0.747\\
$\langle n\rangle \times 10^{15} [cm^{-3}]$&1.39&2.84\\
$\langle T_{i}\rangle [eV]$&23&17\\
$V_{bulk}$ $[km/s]$&20&20\\
\hline
$Computed$&&\\
\hline
$\beta$&0.07&5.5\\
$V_{a} [km/s]$&309&30\\
$C_{s} [km/s]$&31&31\\
$\rho_{i} [cm]$&0.09&0.56\\
$\delta_{i} [cm]$&0.61&0.43\\
$\lambda_mfp^{i} [cm]$&0.16&0.05\\
$f_{ci} [MHz]$&8&1.1\\
$\nu_{i} [MHz]$&6&19\\
$f_{\delta i} [MHz]$&3&4.3\\
$f_{\rho i} [MHz]$&20&3.3\\
\hline
\end{tabular}
\end{table}

Each discharge of the plasma gun constitutes an experimental shot and typically lasts for about 120$\mu s$ from formation of the magnetic fields to their eventual resistive dissipation. The turbulence data reported here is extracted from a time range of 40 to 60$\mu s$. This is the epoch during each shot where the fluctuations are most stationary; energy at the probe location is balanced between injection energy from the gun and loss through advection away from the probe and through possible dissipation mechanisms (including resistive decay of the currents). An ensemble average for each helicity state is constructed from 40 shots.

\section{Analysis Techniques}\label{sec:analysis}

The bulk of the data of this paper is analyzed in frequency space using either a wI avelet transform method~\cite{torrence98} or a traditional Fast-Fourier Transform (FFT). Timeseries data is transformed using a sixth-order Morlet mother wavelet with 256 logarithm scale steps per octave. For magnetic fluctuations, the wavelet transform is applied directly to the $\dot{B}_{j}(t)$ data from the pickup coils, separately for each orthogonal component, $j = r,\theta,z$. Each shot contains 8192 timesteps and the transform is applied to the entire shot to yield an $W_{j}(f,t)$. A known conversion factor is applied to convert the wavelet scales into Fourier frequencies~\cite{torrence98}. The transforms are divided through by the square of the frequency to convert $\dot{B}(f)$ to $B(f)$. Then, the transform is summed over the time range of interest, typically 40-60$\mu s$, yielding,
\begin{equation}
B_{j}(f) = \frac{1}{f^{2}}\sum_{t=t_{0}}^{t_{1}} W_{j}(f,t).
\label{eq:wavelet_transform}
\end{equation}
Since the entire shot is being used in the wavelet transform, the frequency range can extend beyond that which would be typical for an FFT given the particular time range under analysis. The entire frequency range is displayed in the following plots, but the focus of the analysis will be on the same frequency range accessible by an FFT. Details on the comparison between the two approaches is given in Appendix~\ref{sec:WaveFFT}

A similar procedure is applied to the density timeseries from the interferometer and the Mach number timeseries from the Mach probe, but without the frequency scaling used for the magnetic data. For spatial spectra (using separate radial points on the magnetics probe) an FFT is used.

The method for conversion of the magnetic fluctuation spectra from $\vec{B} = (B_{r},B_{\theta},B_{z})$ to $\vec{B} = (B_{\parallel},B_{\perp})$ for the variance anisotropy analysis utilizes the temporal resolution afforded by the wavelet transformation. A local $\vec{B}(t)$ is determined at each time step, $t$, and the local power spectra for each direction, $B_{j}(f,t)$, is projected onto this magnetic field vector to determine the amount of power perpendicular versus parallel. The projection method is detailed in Appendix~\ref{sec:projection2} where this method is also compared to other approaches for determining the anisotropy.

\section{Variance Anisotropy}\label{sec:variance}

\begin{figure}[!htbp]
\centerline{
\includegraphics[width=8.5cm]{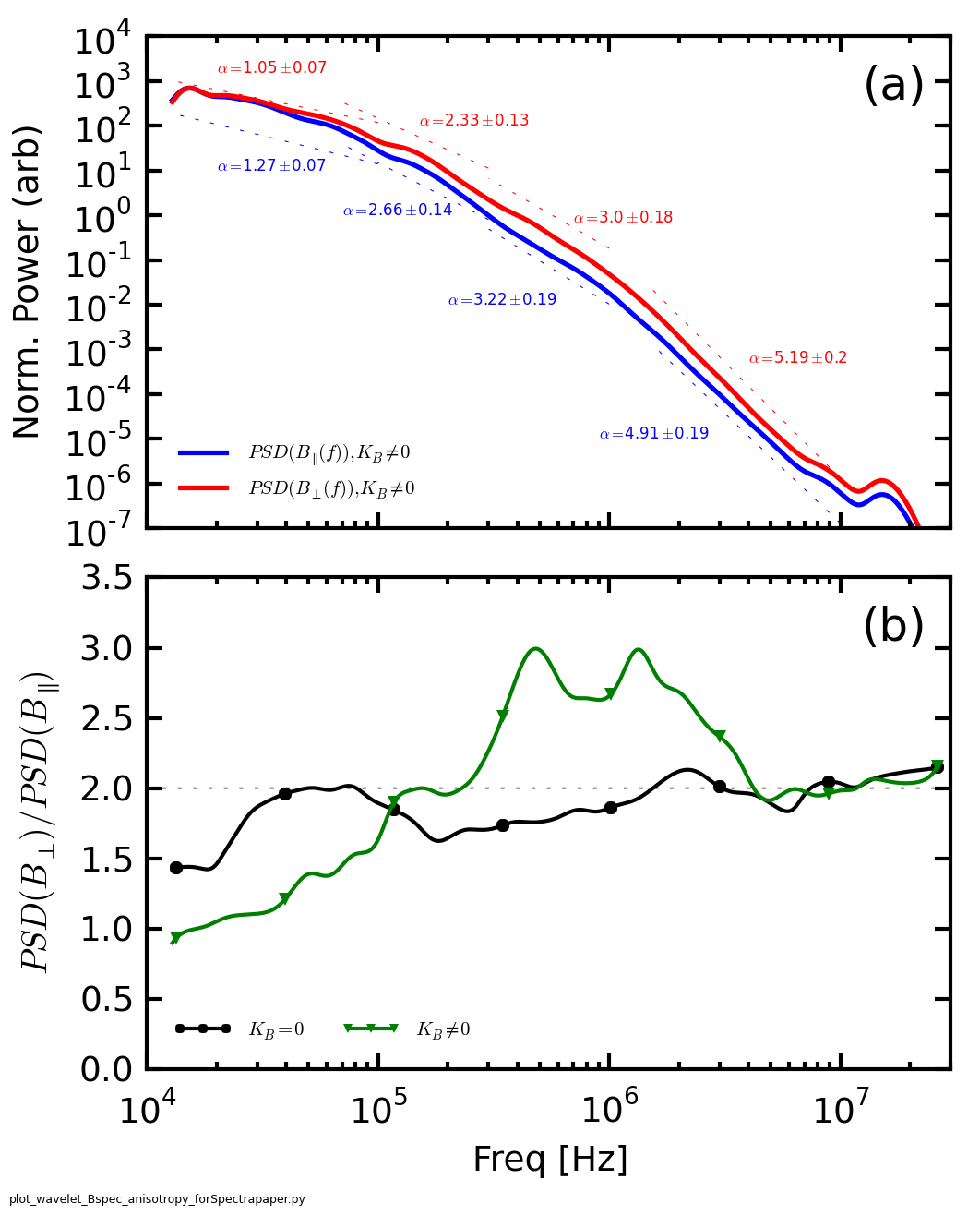}}
\caption{\label{fig:spectra} (a) Magnetic power spectra of the non-zero helicity state portioned into perpendicular (red,higher) and parallel (blue,lower) components with respect to the local magnetic field vector direction. Power law fits and error are indicated for four sections of the spectra. Values and fit regions are indicated in Table~\ref{tab:Bindices}. (b)Ratio of the perpendicular over parallel power spectra from (a) as a function of frequency, plotted on a log-linear scale. The green curve (with triangles) shows the $K_{B}\neq 0$ state while black curve (with circles) shows the result of the same analysis for the zero helicity state. An isotropic ratio is indicated by the horizontal dotted gray line.}
\end{figure}

The magnetic field fluctuation spectra perpendicular and parallel to the local magnetic field vector is shown in figure~\ref{fig:spectra}(a) averaged over 40 shots and the inner four probe tips (a spatial range of 1.5cm). Like previously reported magnetic spectra~\cite{schaffner14a}, both perpendicular and parallel curves exhibit power-law like behavior for most frequencies between 10kHz and 10MHz. Power-law fits to $f^{-\alpha}$ are made to various sections of the curve using a Maximum Likelihood Estimation method~\cite{clauset09} to help compensate for the small ranges of the fits~\cite{deWit13}. The short fits show that the scaling gradually changes from spectral indices of about 1 in the injection (or outer) range to about 5 in the highest frequency sections. As has been noted before, with the exception of the injection range slope, the inertial and possibly dissipation range spectral indices are steeper than observed in solar wind turbulence spectra.

The separation between perpendicular and parallel spectra, however, proceed in a manner similar to that found in space. Both perpendicular fluctuations (red) and parallel fluctuations (blue) begin at roughly the same magnitude in power. Beyond 20kHz, though, the parallel curve dips down slightly faster than the perpendicular and this trend continues up to about 500kHz. Then the separation begins to decrease and approaches a constant value from 5MHz to the Nyquist limit of 32MHz. This gradual change as a function of frequency is more clearly observed in figure ~\ref{fig:spectra}(b) plotted in log-linear format, which provides the ratio of the two curves in figure~\ref{fig:spectra}(a). Since the fluctuations perpendicular to the B-field have two component directions while parallel fluctuations have only one, isotropy---equal fluctuation power in all three components---occurs when the ratio of perpendicular to parallel is two. Isotropy is indicated in figure~\ref{fig:spectra}(b) by the dashed gray line.

At the lowest frequencies, the balance of power actually tips toward parallel over perpendicular. As the frequency increases, the ratio approaches isotropy, then changes over to dominantly perpendicular power. The ratio continues to steadily increase reaching a peak of about 3 at about 500kHz. There is a brief dip before the ratio rises again at about 1.5MHz; at this point, the ratio begins to drop steadily, and approaches an isotropic level at about 5MHz. 

The scale dependency of the $K_{B}\neq 0$ data can be compared to the zero-helicity state. The black curve shows the anisotropy ratio for the zero helicity state, where the field strength is about an order of magnitude less, and there is much less structure to the field. As might be expected for such a state, which has a larger $\beta$~\cite{smith06,sarkar14}, there appears to be very little anisotropy at any scale with values ranging close to R=2.

\begin{figure}[!htbp]
\centerline{
\includegraphics[width=8.5cm]{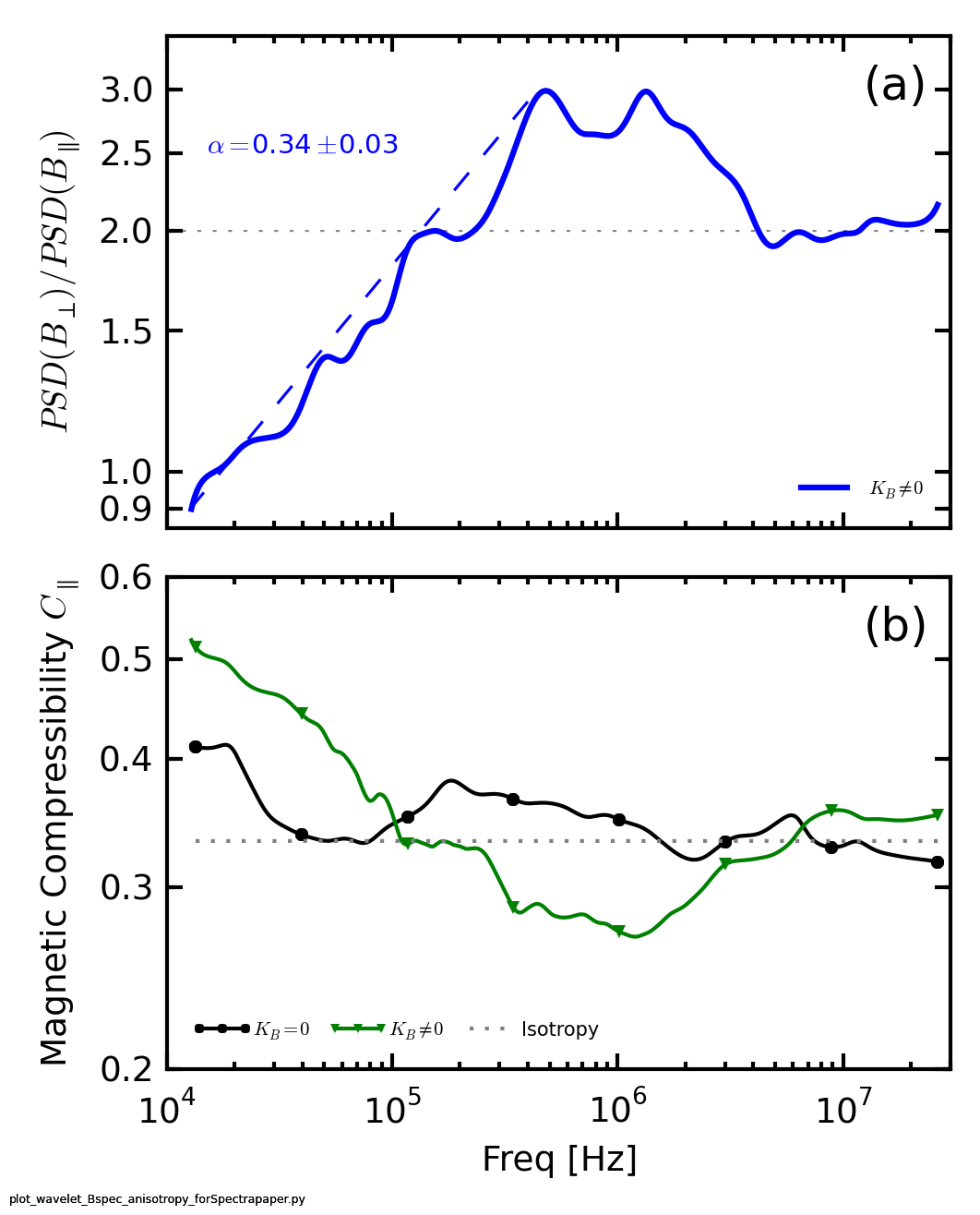}}
\caption{\label{fig:fitratio} (a)The same quantity as shown with the green curve in Figure~\ref{fig:spectra}(b), but now plotted in a log-log format to highlight the power-law behavior as a function of frequency. A fit between 10kHz and 500kHz is made with a spectral index of 0.34 indicated above the curve. (b)Magnetic compressibility, defined in Equation~\ref{eq:magcompress}, for non-zero helicity(green, triangles) and zero helicity(black, circles) as a function of frequency. An isotropic ratio is indicated by the horizontal dotted gray lines in both (a) and (b).}
\end{figure}

\begin{figure}[!htbp]
\centerline{
\includegraphics[width=8.5cm]{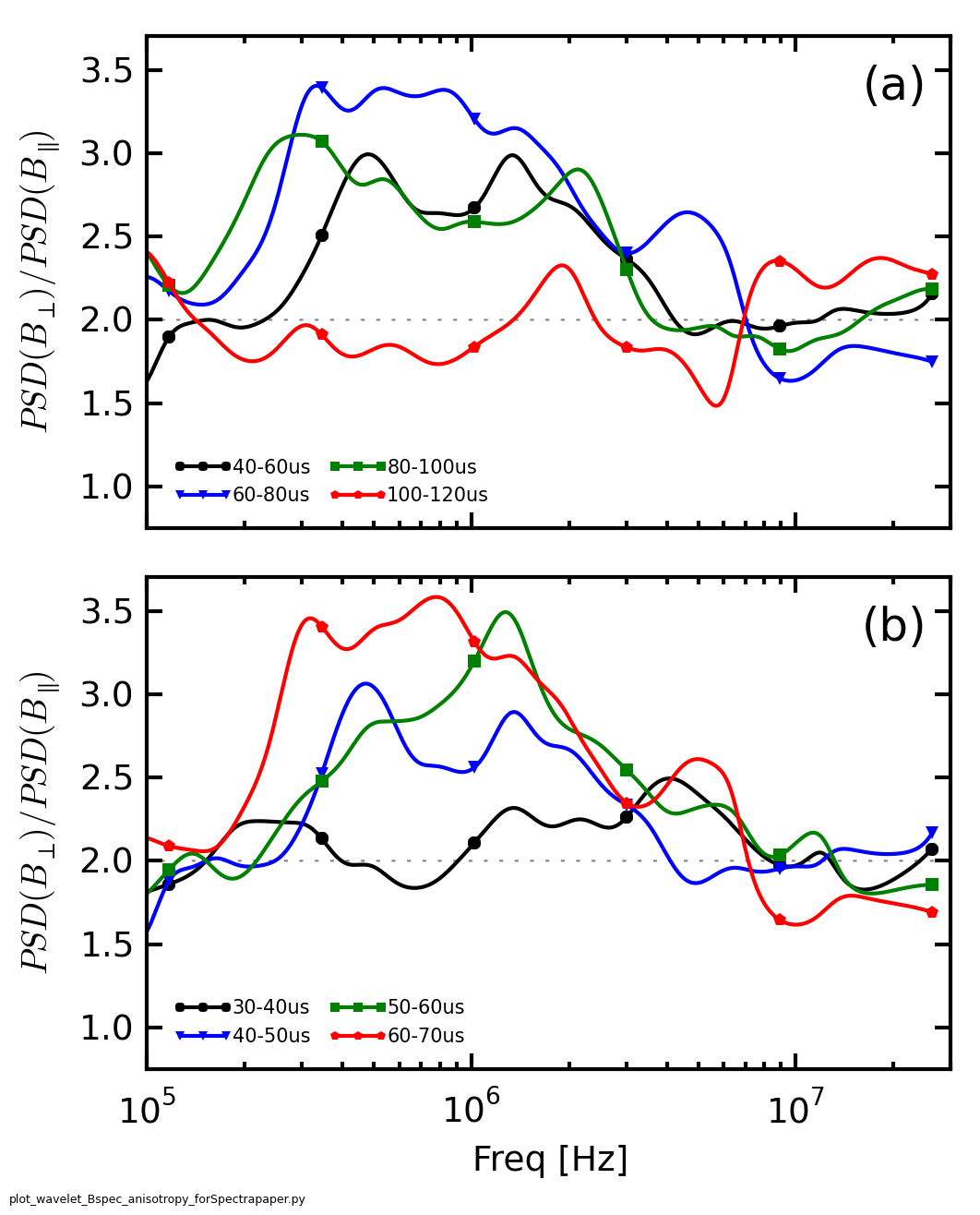}}
\caption{\label{fig:timeratio} (a)Anisotropy ratio for $K_{B}\neq 0$ data for four $20\mu s$ time ranges including and following the nominal analysis range of $40-60\mu s$. (b) Anisotropy ratio for $K_{B}\neq 0$ data for four $10\mu s$ time ranges just before, including, and just after the nominal analysis range. An isotropic ratio is indicated by the horizontal dotted gray lines.}
\end{figure}

Fits to both the spectra and the ratio are shown in figures~\ref{fig:spectra}(a) and figure~\ref{fig:fitratio}(a). In figure~\ref{fig:spectra}(a), fits are made to four sections of each curve. The spectral indices, error and range of fit for both parallel and perpendicular spectra are indicated in the figure and reproduced in Table~\ref{tab:Bindices}. Overall, comparison of the spectral indicies also reflect anisotropy: for each section fit, the parallel slope is consistently steeper than the perpendicular slope. This difference is also reflected in the fit shown in figure~\ref{fig:fitratio}(a). An index of $\alpha = 0.34$ indicates that the ratio scales approximately like $f^{1/3}$ for the region between 10kHz and 500kHz.

\begin{table}
\caption{\label{tab:Bindices}Indices from MLE power-law fits of magnetic fluctuation spectra in Figure~\ref{fig:spectra}.}
\begin{tabular}{cccc}
\toprule
Fit Range[MHz]	&	Direction		&	Index,$\alpha$ for $f^{-\alpha}$	&Error\\
\hline
0.01-0.1				& $\parallel$	& 1.27															&0.07\\
								& $\perp$			& 1.05  														&0.07\\
\hline
0.07-0.3				& $\parallel$	& 2.66															&0.14\\
								& $\perp$			& 2.33  														&0.13\\
\hline
0.3-1.0					& $\parallel$	& 3.22															&0.19\\
								& $\perp$			& 3.00  														&0.18\\
\hline
1.5-10					& $\parallel$	& 4.91															&0.19\\
								& $\perp$			& 5.19  														&0.20\\
\hline
\end{tabular}
\end{table}

An alternate presentation of this ratio is shown in figure~\ref{fig:fitratio}(b), where the ratio of perpendicular and parallel fluctuation power is cast in terms of magnetic compressibility~\cite{kiyani13},
\begin{equation}
C_{\parallel}(f) = \frac{1}{N}\sum^{N}_{i=1}\frac{B_{\parallel}(t_{i},f)}{B_{\parallel}(t_{i},f)+B_{\perp}(t_{i},f)}
\label{eq:magcompress}
\end{equation}
which relates the anisotropy to the characteristic stiffness of the magnetic field structure. Note the ratio between parallel and total is taken before the wavelet spectra are summed over the time range with this definition while the pure ratio shown in previous figures is taken after integration over time.

An evolution of the anisotropy over time is also observed. Figure~\ref{fig:timeratio} shows the change of the anisotropy ratio as a function of frequency for several time ranges of 20$\mu s$ intervals[Fig.~\ref{fig:timeratio}(a)] and 10$\mu s$ intervals[Fig.~\ref{fig:timeratio}(b)]. The black curve in Figure~\ref{fig:timeratio}(b) shows a period of time, 30-40$\mu s$, just as the plasma is reaching the midplane probe. The curve remains near isotropic levels for most of the frequencies. For the later time ranges advancing in 10$\mu s$ intervals, the perpendicular power clearly increases in the 100kHz to 3MHz range. The ratio actually peaks in the time frame just beyond the main analysis period shown in Figure~\ref{fig:spectra}(b). These trends demonstrate that the anisotropy increases as turbulence evolves. Figure~\ref{fig:timeratio}(a), shows that after energy is no longer being injected to maintain the turbulence, the anisotropy decreases. After 100$\mu s$ (red curve), the magnetic field fluctuations have returned to an isotropic ratio.

Some MHD turbulent theories anticipate the ratio of perpendicular fluctuations to parallel fluctuations to increase as a function of decreasing scale size at a particular rate~\cite{goldreich95,boldyrev06}. The scaling of $f^{0.34}$ observed in Figure~\ref{fig:fitratio}(a) compares very well to the theoretical prediction of $B_{\perp}(k)/B_{\parallel}(k) \sim k^{1/3}$ but conditionally on the validity of the Taylor Hypothesis. The temporal evolution in Figure~\ref{fig:timeratio}(b) also suggests an increase in anistropy as the turbulence is given more time to develop. The overall change in anisotropy as a function of frequency (or conversely, the change in magnetic compressibility) can be compared to theories which connect back to the type of fluctuations that may be present in the plasma~\cite{tenbarge12,kiyani13}. Such theories also motivate the comparison between different parameters (such as magnetic and density fluctuations) which can further illuminate the nature of the fluctuations~\cite{klein12} and will be discussed in the following sections.

\section{Flow and Density Spectra}\label{sec:flowdens}

\begin{figure}[!htbp]
\centerline{
\includegraphics[width=8.5cm]{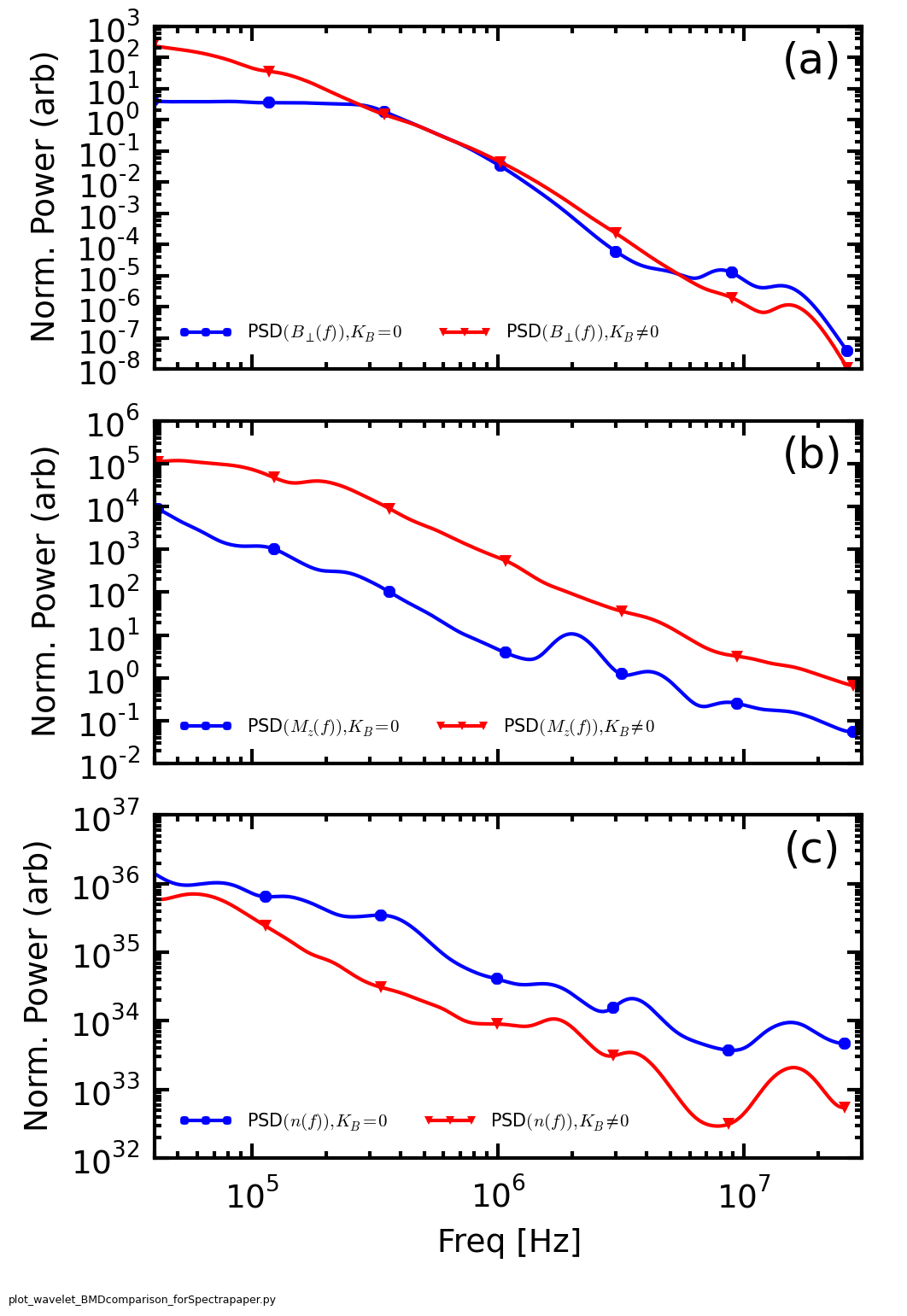}}
\caption{\label{fig:BvsFlow} Comparison of spectra between the non-zero helicity($K_{B}\neq 0$,red triangles) and zero helicity($K_{B}=0$,blue circles) for (a) magnetic power spectra, (b) Mach number power spectra (as a proxy for velocity spectra) and (c) line-integrated density power spectra. Within each sub-plot, the level of each curve is reflective of the relative fluctuation energy difference between each helicity state.}
\end{figure}

For a turbulence cascade to develop, a system needs both energy injection and energy dissipation. The separation of spatial scale between injection and dissipation determines the size of the inertial range. While the actual injection mechanisms of the solar wind are not completely understood, there is evidence from the comparison of large scale magnetic field and velocity fluctuation data that velocity fluctuation energy is being tapped by magnetic fluctuations to sustain an injection-range like cascade for magnetic spectra~\cite{roberts10}. In the SSX plasma, however, the injection scale energy is primarily magnetic---the formation of the unstable spheromaks. This is borne out by similar comparisons of magnetic spectra and velocity spectra. Figure~\ref{fig:BvsFlow} shows magnetic spectra (a) and Mach number fluctuation spectra (b) for the two helicity states. As indicated in Table~\ref{tab:params}, the non-zero and zero helicity states also have different magnetization levels---5kG and 0.8kG, respectively.

Comparison of the blue and red curves in Figure~\ref{fig:BvsFlow}(a) show that the larger mean magnetic fields of the non-zero helicity state are correlated with power in low frequency fluctuations while the lower magnetized state has a flat spectrum at these lower frequiences. Comparing these states with the respective velocity fluctuations seems to indicate that the additional magnetic energy injection at larger scales affects the velocity fluctuation power. For the low-field/zero-helicity state, the velocity cascade is immediately steep as well as lower in energy overall. In the high-field/non-zero helicity state, however, the velocity fluctuation scaling is shallower, but then has a breakpoint around 200kHz where B-field fluctuation energy is about the same for the two helicity states. Furthermore, the higher magnetic energy of the $K_{B}\neq 0$ correlates with higher overall flow fluctuation power of about an order of magnitude. These results suggest that the state with more injected magnetic energy delivers some of its energy to the velocity fluctuations, though more direct evidence of this connection is still being sought.

\begin{figure}[!htbp]
\centerline{
\includegraphics[width=8.5cm]{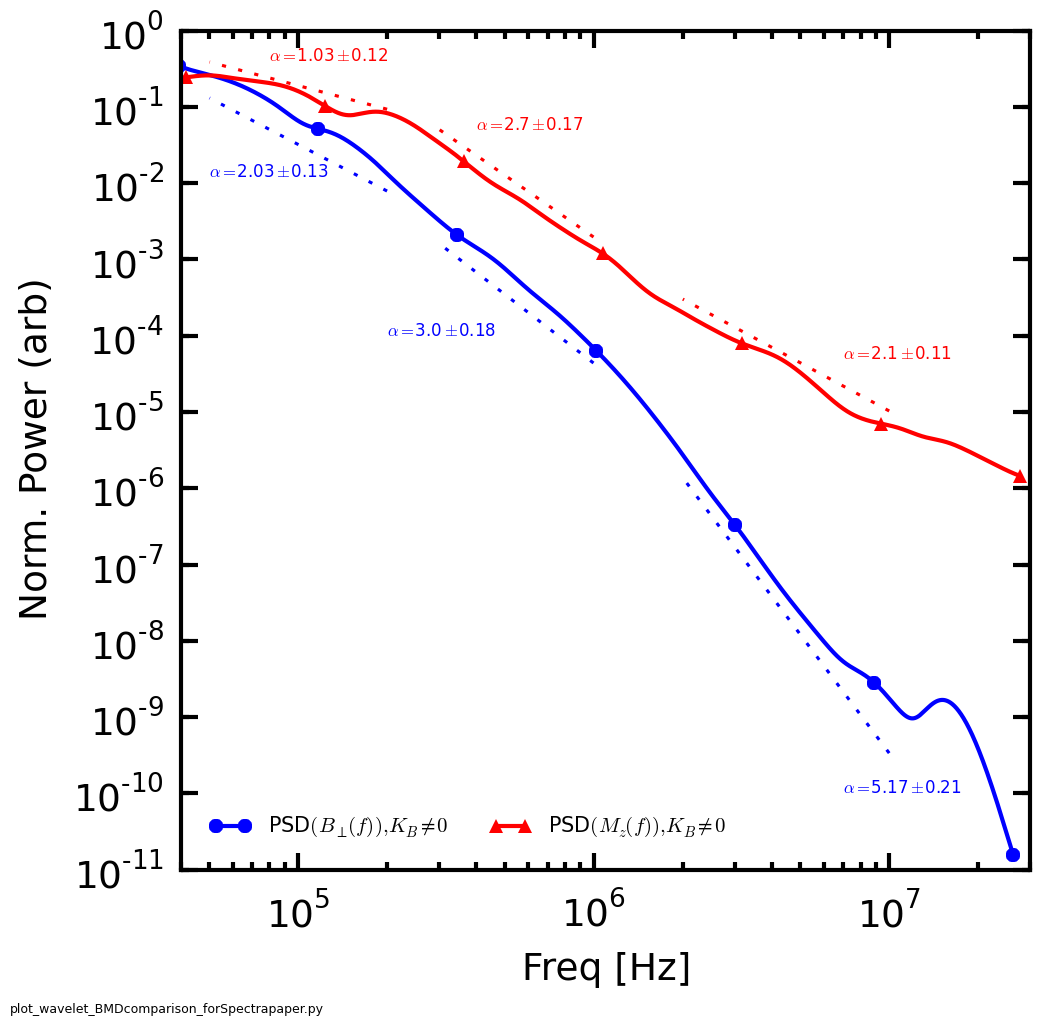}}
\caption{\label{fig:BvsFlow_wFits} Comparison of magnetic power spectra (blue circles) and Mach number power spectra (red triangles) for the non-zero helicity state as a function of frequency. The curves are normalized to their peak values to highlight the comparison of their shapes so relative fluctuation power between the two is not indicated. Fits are made and their values and ranges are indicated in Table~\ref{tab:BMindices}.}
\end{figure}

Figure~\ref{fig:BvsFlow_wFits} shows non-zero helicity state magnetic and velocity fluctuation spectra on the same relative scale with power-law fits indicated in the figure and in Table~\ref{tab:BMindices}. The velocity spectra scales like $f^{-1}$ to about 200kHz, a scaling which is typical for injection range turbulence. Meanwhile, the magnetic spectra is steeper, which implies that some of the magnetic energy may be going to drive flows, though the exact mechanism of this energy transfer is not known and its study is reserved for later work.

\begin{table}
\caption{\label{tab:BMindices}Indices from MLE power-law fits of magnetic and Mach number fluctuation spectra in Figure~\ref{fig:BvsFlow_wFits}.}
\begin{tabular}{cccc}
\toprule
Fit Range[MHz]	&	Parameter		&	Index,$\alpha$ for $f^{-\alpha}$	&Error\\
\hline
0.05-0.2				& $M_{z}$			& 1.03	&0.12\\
								& $B_{\perp}$	& 2.03  &0.13\\
\hline
0.3-1.0					& $M_{z}$			& 2.70	&0.17\\
								& $B_{\perp}$	& 3.00  &0.18\\
\hline
2.0-10					& $M_{z}$			& 2.10	&0.11\\
								& $B_{\perp}$	& 5.17  &0.21\\
\hline
\end{tabular}
\end{table}

Beyond 200kHz, the velocity fluctuations scaling steepens suggesting an inertial range scale. The B-field steepens further as well. Then, the curves match slopes for about a decade in frequency space. Beyond, 2MHz, the B-field scaling drops off significantly while the velocity spectra scaling actually slightly flattens. This is possibly due to a dissipation mechanism that may be further tapping magnetic energy or may be the result of compressive effects~\cite{roberts10} of the plasma. Further analysis must be considered before firm conclusions are made as the spectral signal in this dataset is approaching the bit-depth limit of the Mach probe digitizer. Alternative velocity fluctuation measurements are also being considered (e.g. electric field fluctuations) for future experimental campaigns.

A comparison of density fluctuation spectra from the line-integrated interferometer timeseries is also shown in Figure~\ref{fig:BvsFlow}(c). Note that the fluctuation power of the $K_{B}=0$ state is higher in this case. The spectra curves of the density tend to be the least power-law like, so deriving any conclusions from the data at this point is difficult. There is some evidence for a brief rise in the spectra corresponding to ion inertial length scales (about 1-3MHz) as discussed further in Section~\ref{sec:ionscale}. Future work will attempt to make a more localized measurement as well as seek correlations between magnetic and density fluctuations which can have implications for the underlying nature of the turbulence~\cite{klein12}.

\section{Wavenumber Spectra}\label{sec:wavenumber}

A unique turbulence measurement that can be made in the SSX plasma is a direct wavenumber spectrum using a multi-channel magnetic probe that is inserted radially into the wind-tunnel. The probe can measure $\vec{B}(t)$ at 16 locations along a 7.3cm length of the radius at a spacing of 0.46cm. In Fourier space, this allows measurements of scales from about 7cm to 1cm. Given that the injection scale of the magnetic energy is on the order of the initial size of the spheromaks---15.5cm---and a dissipation scale can be estimated to be just under 1cm---ion inertial length for a 1.5$\times 10^{15}$cm$^{-3}$ plasma is on the order of 0.6cm---the spatial range sampled by the probe is most likely in the inertial range of the turbulent cascade. Note that $k$ and $1/\lambda$ (wavenumber and inverse length) are used interchangeably here, with both intended to mean inverse length scale (i.e. the factor of $2\pi$ is dropped).

Since the probe can take simultaneous measurements of $\vec{B}$ across the plasma, snapshots of the spatial structure of the plasma can be made at each time step. In turn, these spatial distributions can be Fourier transformed to produce power-spectra of the scales. Thus, this measurement can capture the direct wavenumber spectra of the plasma turbulence without reliance on any Doppler shifting as is needed to invoke the Taylor Hypothesis. Moreover, since $\vec{B}$ is constructed from three orthogonal measurements, the power-spectra of vectors perpendicular and parallel to the axial flow of the plasma can be separately analyzed. 

The downside of the measurement of the wavenumber spectra in this way is the lack of resolution compared to a Doppler-shifted frequency spectrum. With only 16 spatial points, the Fourier spectrum can have only 8 points, and only seven can be displayed in log-log format. However, it is a direct measurement and can be used to cross-reference other observations of spectra.

\begin{figure}[!htbp]
\centerline{
\includegraphics[width=8.5cm]{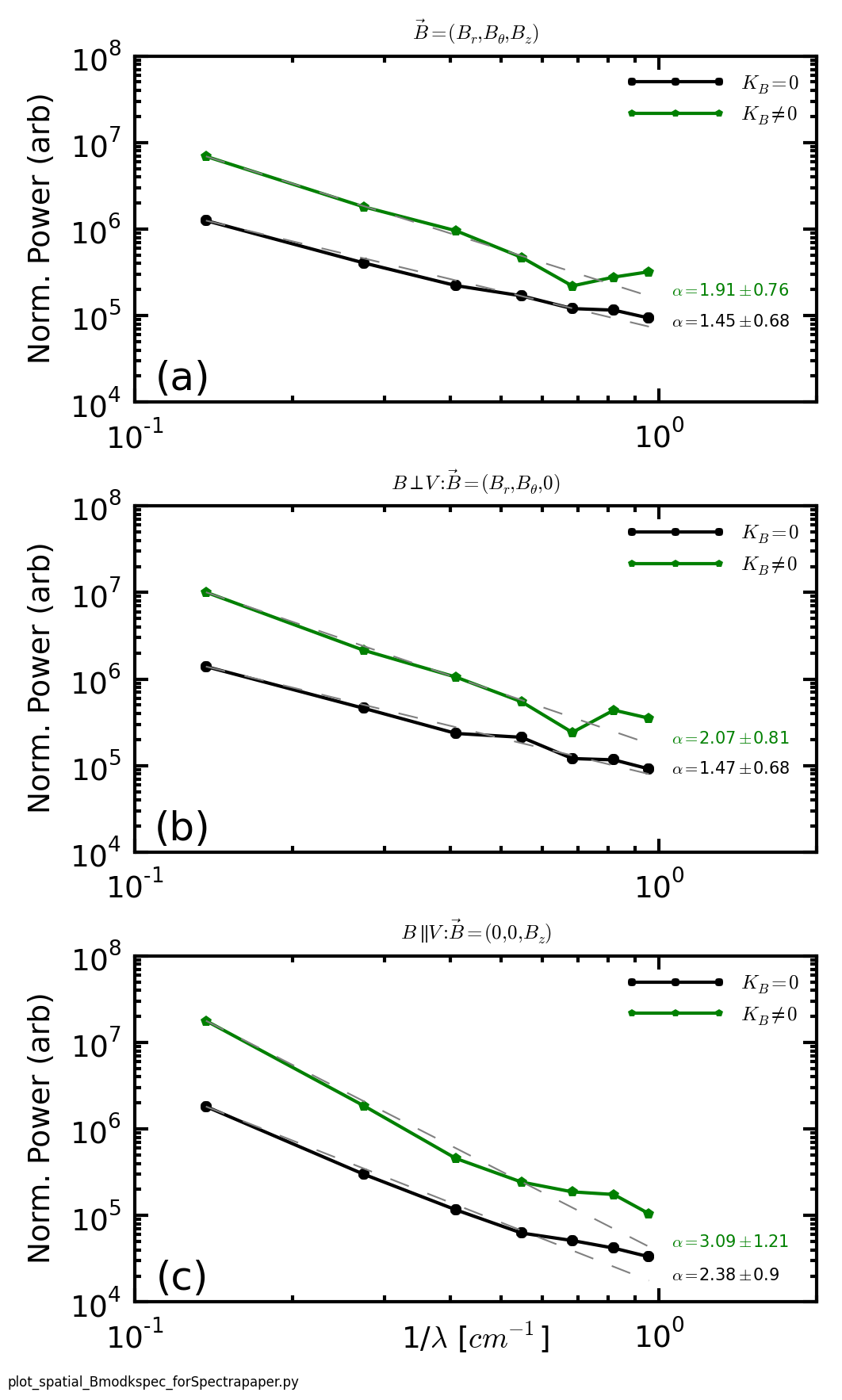}}
\caption{\label{fig:wavenumber_spectra} Magnetic wavenumber power spectra constructed using a multi-channel radially oriented probe for both $K_{B}\neq$ (green,pentagons) and $K_{B}=0$ (black,circles) helicity states. The three panels show this spectrum for (a) the magnitude of the full magnetic field vector, $\vec{B} = (B_{r},B_{\theta},B_{z})$, (b) the magnitude of the projected vector perpendicular to the axial flow, $\vec{B} = (B_{r},B_{\theta},0)$, and (c) the magitude of the vector parallel to the axial flow, $\vec{B} = (0,0,B_{z})$. Power-law fits and errors are indicated at the right side of each curve and are fit over the entire data range shown.}
\end{figure}

Figure~\ref{fig:wavenumber_spectra} shows the wavenumber power spectrum for the two helicity states and for full-vector(a), perpendicular vector(b) and parallel vector(c), averaged for each timestep between 40 and 60$\mu s$ and over 40 shots. Comparison of the curves in Figure~\ref{fig:wavenumber_spectra}(a) seem to show a slight variation in the slopes between helicity states, though with the large given errors in the fit (due to low resolution), the slopes of the two curves are essentially the same. A similar trend is observed between the curves in (b) and (c) as well. A larger difference arises when comparing the curves of different vectors. Namely, the separately computed perpendicular and parallel spectra in (b) and (c) tend to be slightly steeper than the full vector spectra in (a). Moreover, the parallel curves appear to be steeper than the perpendicular curves. Again, the error in the fits are large for the low resolution data, but the trends are very suggestive of a wavenumber anisotropy (rather than variance anisotropy). However, there is not as much conclusive evidence for this form of anisotropy as there is for the variance anisotropy.

\begin{figure}[!htbp]
\centerline{
\includegraphics[width=8.5cm]{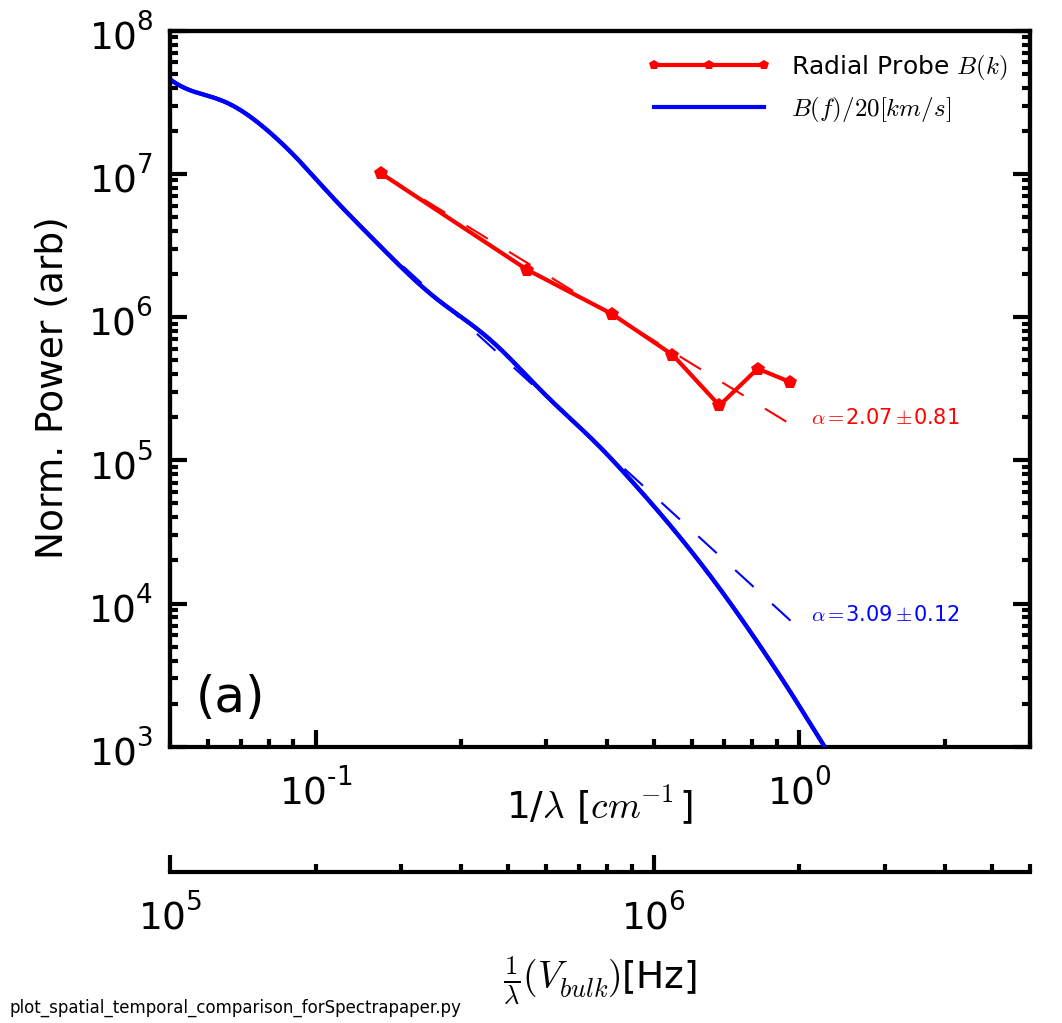}}
\caption{\label{fig:wavenumber_comp} Comparison of a direct wavenumber spectrum (red, pentagons) and a frequency spectrum (blue) shifted into spatial scales using the Taylor Hypothesis and a bulk flow of 20km/s as a function of inverse length scale. Fits are made to the same range in inverse length space and are indicated to the right of each curve. The double x-axes show the relative correspondence between inverse length scales and frequencies given the Taylor Hypothesis shift.}
\end{figure}

The wavenumber spectra and frequency spectra can be directly compared by invoking the Taylor hypothesis for the frequency spectra and Doppler-shifting the frequency spectra by the bulk plasma velocity,$V_{p}$,
\begin{equation}
B(f) \longrightarrow B(f-kV_{p}) \longrightarrow V_{p}B(k).
\label{eq:Taylor_Hyp}
\end{equation}

Typically, the last simplification in Equation~\ref{eq:Taylor_Hyp} can only be done if the bulk flow is high enough such that $kV_{p} \gg f$, where $f$ is the temporal frequency of any modes in the plasma. However, in this plasma, the nature of the present modes is still under investigation so for a first approximation comparison, the full Taylor Hypothesis transformation is used. For this plasma, the bulk velocity can be estimated (using both time-of-flight and Mach probe measurements) to be about 20km/s. The frequency spectra can be plotted on the same scale as the wavenumber spectra. Figure~\ref{fig:wavenumber_comp} shows this comparison for the non-zero helicity plasma. The bottom axes show how use of the Taylor Hypothesis relates scales of 10cm-1cm to frequencies of 200kHz-2MHz. The curves are placed arbitrarily on the y-axis. Maximum likelihood estimation power-law fits are made to the same range in both curves. The slopes of the curves are comparable suggesting that invoking the Taylor hypothesis for the frequency spectra is not entirely unwarranted. Instead, the steeper slope of the frequency spectra could be reflective of the effect of a combined temporal {\it and} spatial scaling, which the direct wavenumber spectrum does not include. However, breakdown of Taylor Hypothesis has been theoretically predicted to make the spectra more shallow than steeper(Klein Taylor Hypothesis draft). It also should be noted that these two types of spectra are separately sampling the radial and axial wavenumber spectra. If the turbulence were completely isotropic, one would not expect a difference between the wavenumber spectra. Thus, the differences observed might also be reflective of a wavenumber anisotropy as the direct wavenumber spectra probes $k_{r}$ and the Doppler-shifted frequency spectra probes $k_{z}$. This possible anisotropy is also hinted at in Figure~\ref{fig:wavenumber_spectra}. Future experiments will seek to make a more direct comparison by aligning a multi-channel probe in line with, rather than perpendicular to, the flow, in order to further test the validity of the Taylor Hypothesis in this case.

\section{Comparison with Simulation}\label{sec:simulation}

Simulations of the plasma produced in the SSX wind-tunnel have been conducted within the HiFi spectral-element multi-fluid modeling framework using a set of normalized compressible resitive and Hall-MHD equations. Computational details of the simulations analyzed here have been previously reported~\cite{schaffner14a} and shown to have favorable comparisons of turbulent spectra and intermittency between simulations and experiment. Further analysis is presented here which shows similar observations of anisotropy, wavenumber spectra, and velocity-B-field spectra comparisons to that observed in the experiment. The simulations have parameters most closely resembling the $K_{B}\neq 0$ helicity state. 

Timeseries of quantities in 3mm spheres approximately 1cm off the central axis and at the midplane are extracted from the simulations for density, three axes of magnetic field, three axes of velocity. Three axis magnetic fluctuations at 24 radial locations at the midplane are also extracted. To provide some ensemble averaging, points at eight different azimuthal angles are used since only one iteration of the simulation discharge with each of the two MHD models, resistive and Hall, is presently available.

\begin{figure}[!htbp]
\centerline{
\includegraphics[width=8.5cm]{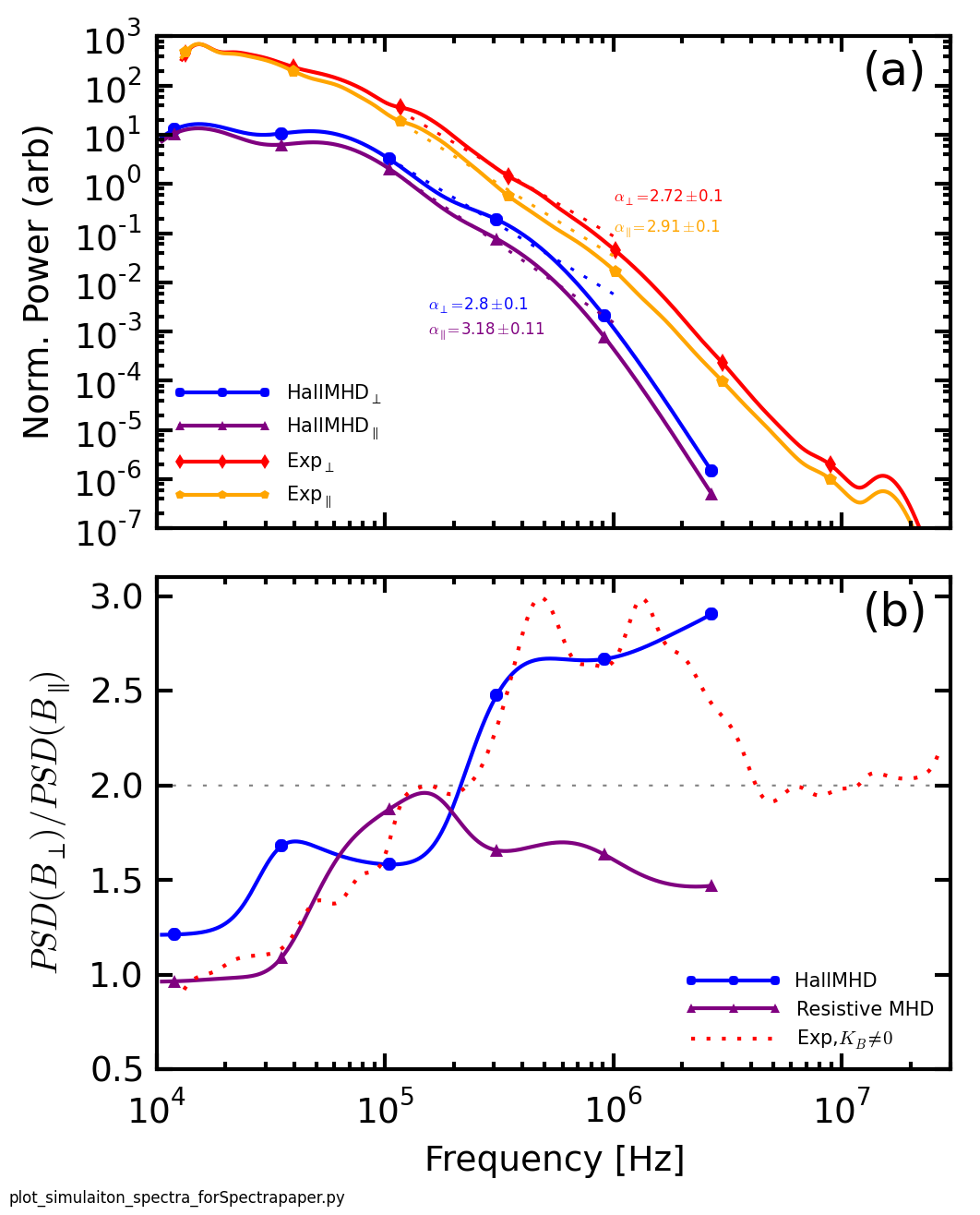}}
\caption{\label{fig:aniso_comp} (a) Perpendiuclar and parallel magnetic power spectra for simulation(blue circles, purple triangles) and the non-zero helicity experiment(red diamonds, orange pentagons) as a function of frequency. The experimental and simulation curves are arbitrarily placed on the y-axis, but the perpendicular(upper) and parallel(lower) curves within each pair are placed relative to each other. Fits and fit regions are indicated in Table~\ref{tab:Simindices}. (b) Anisotropy ratio as a function of frequency for a Hall-MHD simulation(blue circles), a resistive-MHD simulation(purple triangles) and the non-zero helicity experimental data(dotted red). An isotropic ratio is indicated by the horizontal dotted gray line.}
\end{figure}

\begin{table}
\caption{\label{tab:Simindices}Indices from MLE power-law fits of magnetic fluctuation spectra from experiment and simulation in Figure~\ref{fig:aniso_comp}.}
\begin{tabular}{cccc}
\toprule
Fit Range[MHz]	&	Parameter						&	Index,$\alpha$ for $f^{-\alpha}$	&Error\\
\hline
0.1-1.0					& Sim:$B_{\parallel}$	& 3.22	&0.11\\
								& Sim:$B_{\perp}$			& 3.26  &0.11\\
								& Exp:$B_{\parallel}$	& 2.91	&0.10\\
								& Exp:$B_{\perp}$			& 2.72  &0.10\\
\hline
\end{tabular}
\end{table}

The simulation timeseries data is analyzed in a similar manner as the experimental data with the exception that the mother wavelet used for the wavelet transform of the simulation data is a fourth-order Paul rather than a sixth-order Morlet, in order to better capture time resolution for the lesser sampled simulation. A variance anisotropy analysis is conducted in the same manner, as well, using a local magnetic field and the projection method. Figure~\ref{fig:aniso_comp}(a) shows simulation decomposition into perpendicular and parallel spectra compared to $K_{B}\neq 0$ experimental spectra. The simulation data is averaged over eight radially spaced points spanning about 2cm. The simulation and experimental spectra are staggered on the y-axis to emphasize features of the shape, but the perpendicular and parallel curves of each are relative to one another. Clearly, the simulation data exhibits growing variance anisotropy with increasing frequency. The slopes of the simulation spectra match qualitatively well in the region of 100kHz to 1MHz, though the fit spectra indices indicate a slightly steeper slope than the experiment (see Table~\ref{tab:Simindices}). The high frequency end of the simulation spectra drops in power faster than the experiment, likely due to the limits in time resolution.

The trend in anisotropy ratio is also very similar in the simulation and the experiment, as seen in Figure~\ref{fig:aniso_comp}(b). Though the simulation curve does not achieve as large a peak ratio, it does level-off at about the same frequency. The clear observation of an increase in ratio suggests that the compressible Hall-MHD physics captures the generation of the anisotropy. Figure~\ref{fig:aniso_comp}(b) also shows the anisotropy ratio for a simulation run with the Hall term in the compressible MHD equations set to zero (i.e. resistive MHD). Unlike the Hall MHD and the experiment, the ratio does not switch over to perpendicular dominance, and instead stays near or below the isotropy line. The implications of this have not been analyzed in depth though a Hall term is not considered necessary for anisotropy to develop. Also note that the simulation ratio does not decrease again at high frequencies compared to the experiment, but this may be due to the lower frequency resolution of the simulation.

\begin{figure}[!htbp]
\centerline{
\includegraphics[width=8.5cm]{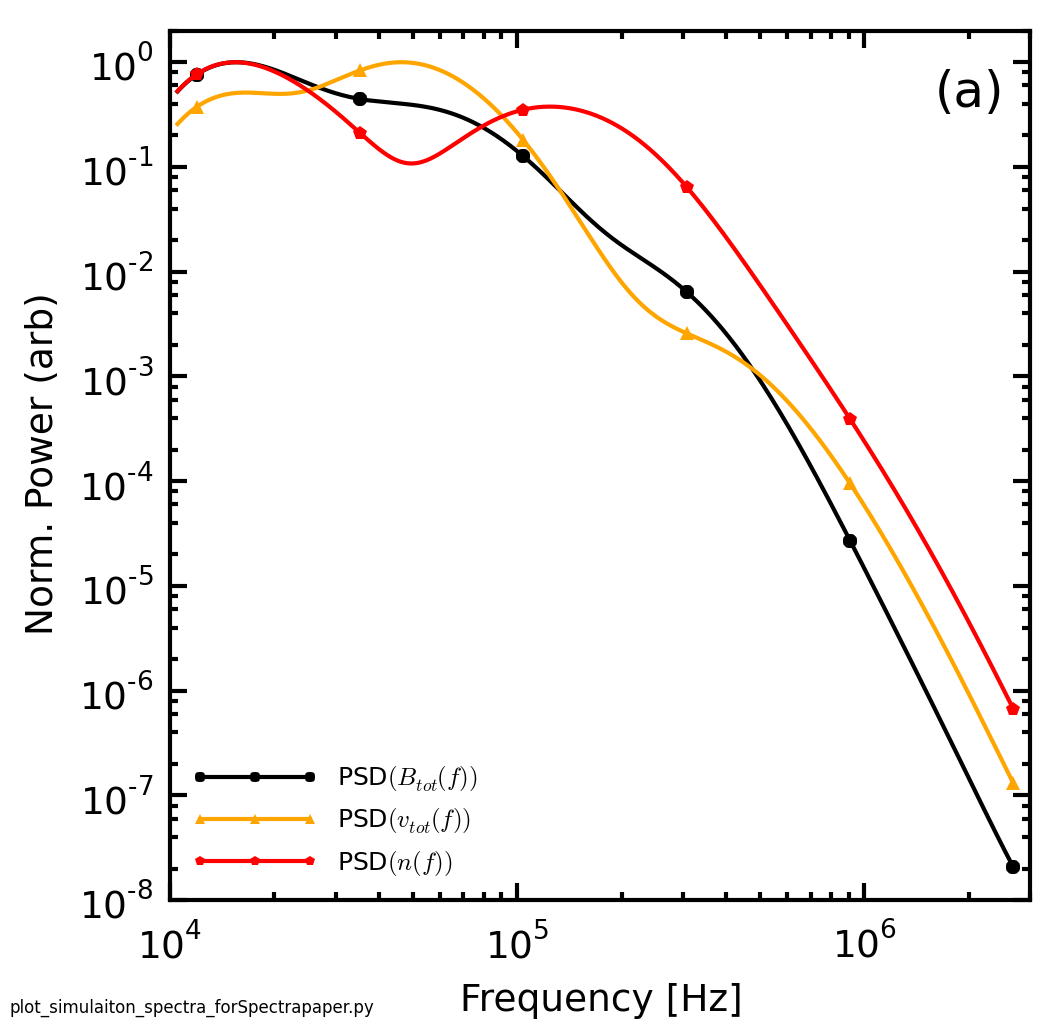}}
\caption{\label{fig:bflow_comp} Comparison of simulation generated magnetic(black circles), velocity(orange triangles) and density(red pentagons) power spectra as a function of frequency.}
\end{figure}

A comparison between velocity and magnetic field fluctuations in the simulation can also be made. Figure~\ref{fig:bflow_comp} shows wavelet transformed frequency power spectra for the total magnetic field (sum of $B_{r}$, $B_{\theta}$, and $B_{z}$), total velocity, and density, all normalized to their respective peaks. Qualitatively, the comparison between velocity and magnetic field spectra supports the results of the experimental data for non-zero helicity plasmas: the peak in the velocity spectra occurs at a larger frequency than the magnetic spectra. This suggests that energy for the velocity fluctuations are being injected at a smaller scale than the magnetic field fluctuations. Though not conducted here, further analysis of the simulation could potentially show direct energy transfer between the magnetic and velocity fluctuations.

\begin{figure}[!htbp]
\centerline{
\includegraphics[width=8.5cm]{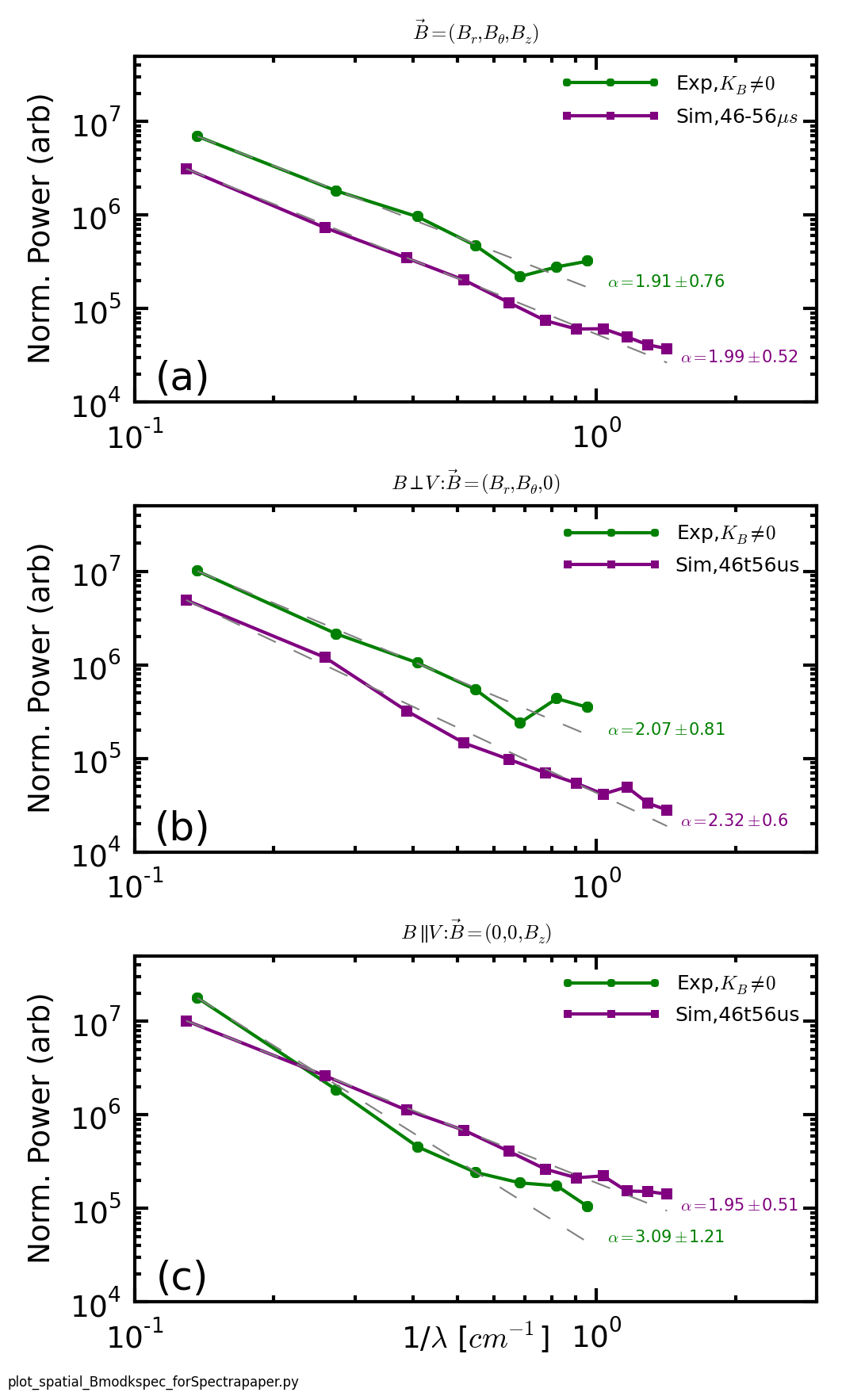}}
\caption{\label{fig:sim_wavenumber_comp} Direct magnetic wavenumber power spectra from the non-zero helicity data and the Hall-MHD simulation. The three sections indicate the same vector choices described in the caption of Figure~\ref{fig:wavenumber_spectra}. Fits to each are displayed to the right of each curve and comprise the entire shown data range for each respective curve. The simulation probes slightly smaller spatial scales since 24 points are used to span the same spatial range rather than 16 points for the experimental probe.}
\end{figure}

\begin{figure}[!htbp]
\centerline{
\includegraphics[width=8.5cm]{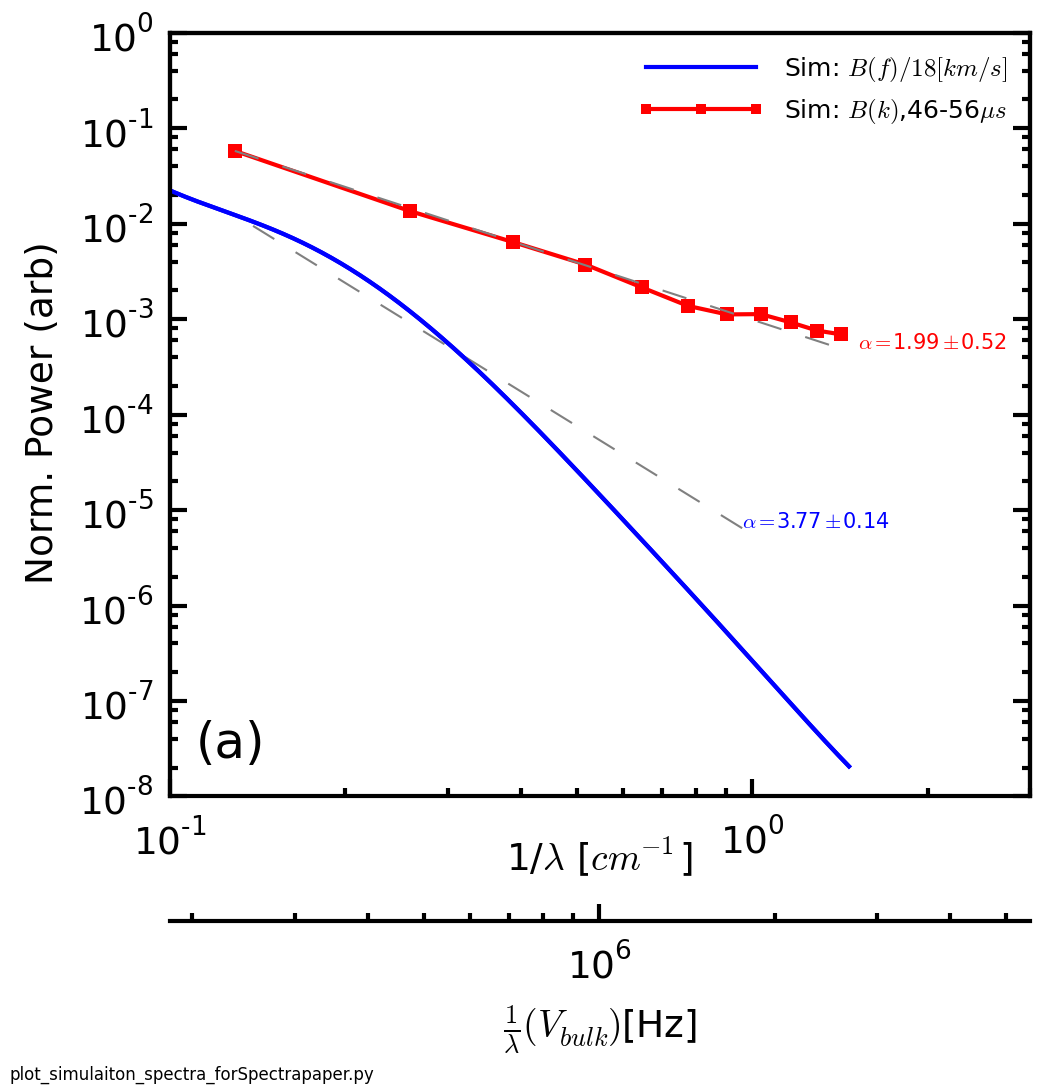}}
\caption{\label{fig:sim_spatial_comp} Similar comparison of wavenumber and converted frequency spectra for simulation data as made in Figure~\ref{fig:wavenumber_comp}, though for the simulation the bulk flow is 18km/s so the correspondence between the two x-axis scales is slightly different than the previous plot.}
\end{figure}

The wavenumber spectra from a radial cut of the simulation data is also generated using 24 points. Figure~\ref{fig:sim_wavenumber_comp} shows a direct comparison between simulation and experimental wavenumber spectra. The slightly higher spatial resolution allows the simulation to reach a smaller scale than the experiment, to about 0.7cm. In general, the comparison between simulation and experiment is good suggesting that the simulation is capturing well the spatial structure of the turbulence. Even though the simulation can observe slightly smaller scales, it does not appear to probe small enough to exhibit any dissipation effects at ion inertial length scales, which for the simulation is at about 0.7cm.

Like in the experiment, the spatial and temporal spectra of the simulation is compared and shown in Figure~\ref{fig:sim_spatial_comp}. The simulation has a bulk axial flow of 18km/s, close to the 20km/s observed in the experiment. A similar trend is seen with the spatial spectra having a shallower slope than the Doppler-shifted frequency spectra. The main difference again appears to be that the frequency spectra hits the limits of the temporal resolution at lower frequencies than the experiment.

These further comparisons of turbulent statistics and characteristics between the experimental plasma and a compressible Hall-MHD simulation help validate the model as useful for understanding the physical processes. Subsequent simulation analysis will likely entail more detailed computation of how energy might be being distributed and moved through the plasma including relationships between magnetic field and velocity as well as between magnetic fluctuations perpendicular and parallel to a local B-field. Higher time resolutions to probe ion scale physics and the effects of varying helicity can also be explored.

\section{Discussion and Evidence for Ion Scale Effects}\label{sec:ionscale}

\begin{figure}[!htbp]
\centerline{
\includegraphics[width=8.5cm]{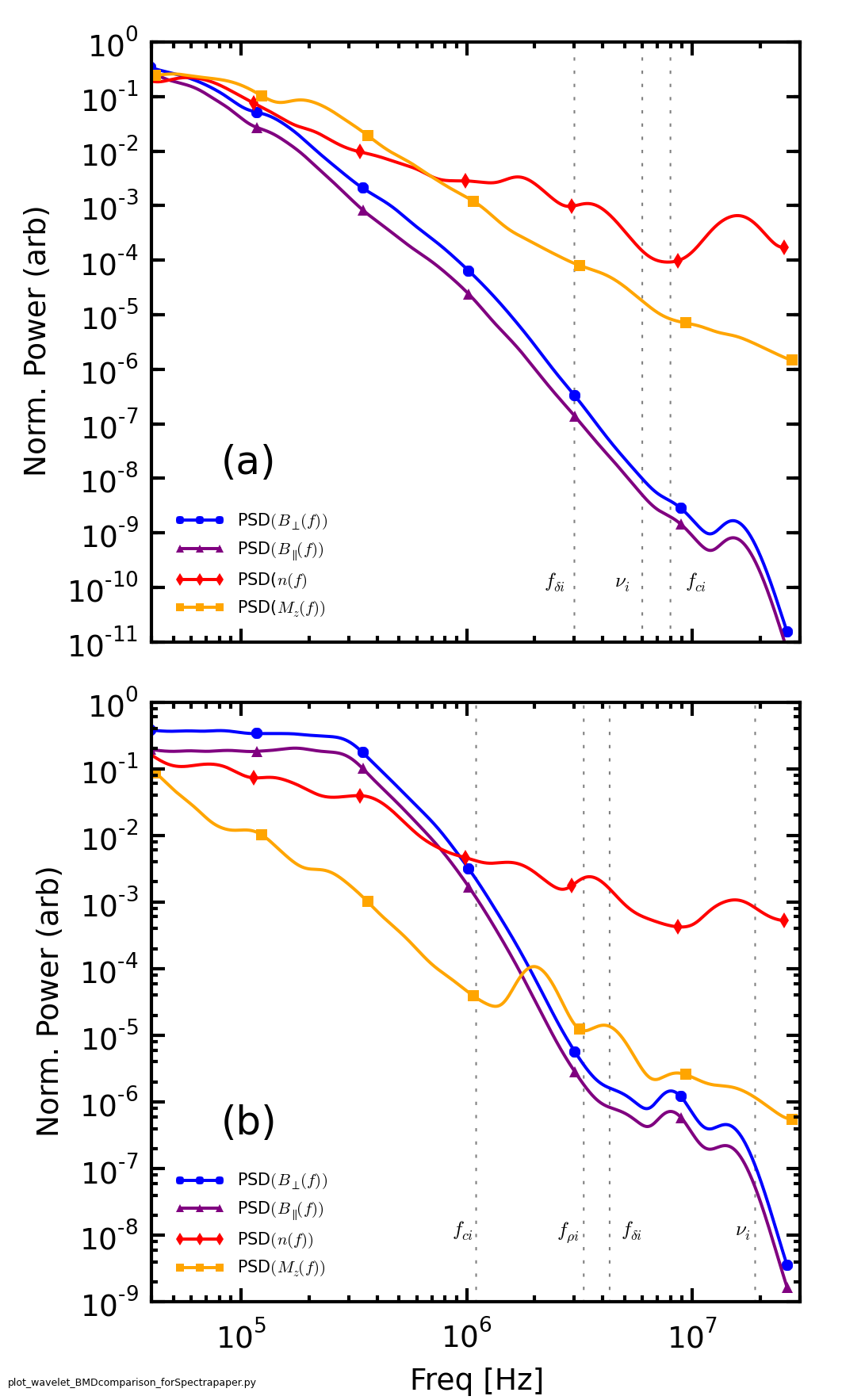}}
\caption{\label{fig:BMD_comp} (a) Comparison of magnetic(perp-blue circles,para-purple triangles), Mach(orange squares), and density(red diamonds) spectra all normalized to their respective maximum value to highlight differences in spectral shape for $K_{B}\neq 0$ data. Vertical dashed lines indicate the position in frequency space of the Doppler-shifted ion inertial length, $f_{\delta i}$, the collision frequency, $\nu_i$, and the ion cyclotron frequency,$f_{ci}$. (b) Similar comparison as (a), but for the zero helicity case. The Doppler-shifted ion gyroradius, $f_{\rho i}$ is also indicated on this second subplot.}
\end{figure}

A major remaining question for this analysis is whether the plasma diagnostics are able to observe effects of a dissipation scale in this turbulence. Perhaps, a more general question can be posed: How well does this plasma exhibit a traditional fluid-turbulence-like picture which posits an injection scale, inertial scale and dissipation scale?

The results of this analysis provide a number of hints that the ion inertial scale length is being probed, but no one piece of evidence is strong enough to make a conclusive assertion. The first clue arises by comparing the break-point of the magnetic spectra with the Doppler-shifted ion inertial scale length, $f_{\delta i}$. Figure~\ref{fig:BMD_comp}(a) shows the spectra for magnetic field, Mach number and density for the $K_{B}\neq 0$ data all normalized to their respective maximum value, with dashed lines indicating the Doppler-shifted frequency of ion inertial length, $f_{\delta i}$, using a bulk flow speed of 20km/s, the collision frequency, $\nu_{i}$, and ion cyclotron frequency, $f_{ci}$. Note that the break point occurs just before the ion inertial frequency is reached. Since the ion inertial scale is often associated with the scale size of reconnection layers or current sheets, a break-point just proceeding this scale suggests the onset of a dissipation mechanism associated with current sheets of some form.

Supporting pieces of evidence for this hypothesis come from the comparison of the density, flow and magnetic spectra, comparison to trends in variance anisotropy, and comparison to observations in space plasmas. The red curve in Figure~\ref{fig:BMD_comp}(a) shows a slight bump up around $f_{\delta i}$. This is possibly evidence of the density bump effect observed when the plasma becomes more compressible---which would be expected at ion inertial length scales. The flattening of the velocity spectrum has also been speculated to be a signal of compressive effects~\cite{roberts10}, but no clear cut relationship has as yet been observed. Similarly, the anisotropy ratio in Figure~\ref{fig:spectra}(b) begins to decrease at about this frequency (see Figure~\ref{fig:fitratio}) which also suggests an increase in compressibility. Such decreases in anisotropy ratio at dissipation scales have been observed in the solar wind~\cite{hamilton08,kiyani13}. The spectral slopes and break-point at $c/\omega_{pi}$ of the magnetic spectrum also compare well with observations made in the magnetosheath~\cite{yordanova08}.

However, there remain other explanations for these effects. One, the Taylor Hypothesis used to establish the connection between the frequency and the scale length here is not, as yet, conclusively applicable. If the Taylor Hypothesis cannot be invoked, there may be other reasons for the break-point observed. Also, the flattening of the spectra in Figure~\ref{fig:BMD_comp}(a) may be due to reaching the noise floor set by the bit-depth of the interferometer/Mach probe data acquisition system (smaller than the 14-bit range of the magnetic). Moreover, there is evidence, at least in the solar wind, that density spectra might be expected to steepen beyond ion scales~\cite{chen12}, not flatten.

Comparison of the two different helicity states only adds to the ambiguity. Figure~\ref{fig:BMD_comp}(b) shows the same curves as in (a), but for the zero-helicity state. The break-point in the magnetic field here appears to occur close to the ion cyclotron frequency rather than the ion inertial length. It should be noted though that previous work~\cite{schaffner14b} has suggested that the zero-helicity state consists of much fewer current sheets and as such, dissipation in this state would depend much less on those mechanisms. The zero-helicity state also shows no change in anisotropy with scale nor any density bump.

The results presented, nevertheless, do highlight the need for further investigation, particularly in whether a Taylor Hypothesis can be invoked, or the very least, whether some type of Doppler shift can be applied to properly connect the frequency and scale size of the signal. Similarly, a higher resolution in the spatial probes could provide confirmation as its current incarnation just misses the apparent dissipation scale. Moreover, better resolved density and Mach flow diagnostics would be useful in order to distinguish between noise and real effect.

Lastly, the simulation could potentially provide some insight into the processes occurring. Unfortunately, the comparison with the current incarnation of the simulation diverges at scales associated with ion physics. This indicates the need to reexamine the simulation at higher temporal sampling rates, but could also reflect the need to include other terms such as ion finite Larmour radius (FLR) effects or gyroviscosity in order to better reproduce the correct dissipation physics.

%Lastly, the simulation could potentially provide some insight into the processes occurring. Unfortunately, the simulation fails to probe at a small enough scale to show ion scale effects. Both the frequency spectra and wavenumber spectra fall short of ion inertial scales. The variance anisotropy ratio supports this as an increase in the ratio is observed, but not a decrease at smaller scales/higher frequencies. However, the favorable comparison between the experiment and the simulation at smaller frequencies and larger scales means that improvement of the simulations resolution maybe be useful for exploring the dissipation physics.

\section{Conclusions}\label{sec:conclusions}

This manuscript presents a broad analysis of both temporal and spatial fluctuation spectra with the intent of understanding the MHD turbulence observed in a wind-tunnel on the Swarthmore Spheromak Experiment. The results show the presence of variance anisotropy---more power in fluctuations perpendicular to the local magnetic field than parallel---for a plasma with non-zero injected helicity. The ratio of perpendicular magnetic fluctuation power to parallel shows variation as a function of frequency and of time, reaching peak values of about R=3 during the period of mostly stationary fluctuations and at frequencies between 500kHz and 1MHz. Very little anisotropy is observed for a zero-helicity state, which may be more closely related to the value of $\beta$ for this state of the plasma rather than the helicity itself.

Comparison of fluctuation spectra of magnetic fields and velocity appear to support the notion that energy is primarily injected into the plasma through magnetic fields (during the spheromak formation process). Comparison amongst magnetic, velocity, and density spectra provide clues for the nature of the dissipation processes occurring in the plasma, but much more research is needed for conclusions to be made.

A direct wavenumber spectrum is also measured in the plasma, highlighting a possible advantage that measurements in laboratory plasmas can have over {\it in-situ} space measurements. The wavenumber spectra show slight differences when compared to temporal frequency spectra converted into spatial scales using the Taylor Hypothesis and the measured bulk axial flow of the plasma. Wavenumber spectra also suggest a wavenumber anisotropy, but the results are less conclusive than the observation of the variance anisotropy. These initial results, however, provide impetus for more detailed spatial turbulent measurements in future experimental campaigns including the possibility of testing the validity of the Taylor Hypothesis in this plasma.

All of these results are also compared to a Hall-MHD computation generated to simulate the spheromak relaxation process. Turbulent spectra using synthetic diagnostics compare favorably to the experiment particularly in the measured spectral indices for power-law fits, but comparisons tend to diverge at higher frequencies where more ion scale effects may be in play. Improved data extraction techniques are being sought in order to produce higher temporally resolved simulation data.

Many open questions still remain though, including whether or not dissipation at the ion inertial length scale is observed. Both the magnetic spectra breakpoint and a decrease in anisotropy occur at scales associated with the ion inertial length, but only under the assumption of Taylor Hypothesis. The direct wavenumber spectra do not probe a small enough scale to provide any evidence for or against this hypothesis. The encouraging comparisons with simulation, though, might provide some support if it can be pushed to higher resolutions. Other diagnostic techniques are also being pursued to help unravel this issue.

The goal of this analysis is to establish the use of the MHD wind-tunnel as a testbed for understanding turbulence typically researched at space physics scales. Many of these results have intriguing comparisons with turbulent situations in the solar wind and the magnetosheath---both for cases where the data compares well, as well as where it does not. The advantage of laboratory experiments lies in the ability to more easily make spatial measurements, the ability to have some nominal control over parameters, and the ability to make many repeatable measurements. Through a combination of these laboratory observations, computer simulation {\it and} space observation, a greater understanding of MHD turbulence is sought.

\appendix

\section{Computation of Variance Anisotropy}\label{sec:projection}

The computation of the variance anisotropy involves two main steps: a determination of fluctuation power and an estimate of the distribution of the fluctuation power relative to a vector direction. The first step is accomplished using a wavelet transform procedure as is discussed in the text. The division of fluctuation power is accomplished using what will be described here as a projection method. A second estimate of fluctuation power distribution is constructed using a threshold method and is also described in detail in this appendix. The threshold method is a more straightforward process for determining the level of variance anisotropy, but suffers from diminishing statistics. Its presentation here is mainly as a validation of the eventual use of the projection method for the analysis in the paper. 

\subsection{Threshold Method}\label{sec:threshold}

Both the threshold method and the projection method for determining variance anisotropy rely on the ability of the wavelet transform to yield a power spectrum distribution (PSD) as a function of both time and frequency (i.e. B(f,t)). A local (in time) vector $\vec{B}$ is determined for each time, t,
\begin{equation}
\vec{B}(t) = B_{r}(t)\hat{r} + B_{\theta}(t)\hat{\theta} + B_{z}(t)\hat{z}
\label{eq:Bvector}
\end{equation}
where each component $j=r,\theta,z$ is determined from $\dot{B}_{j}$ by integrating over time as
\begin{equation}
B_{j}(t) = \int_{0}^{t} \frac{d}{d\tau}\left(B_{j}(\tau)\right) d\tau.
\label{eq:Bintegrated}
\end{equation}

Since the magnetic probe measures orthogonal magnetic field directions by construction, this fact can be used to directly seek a difference in fluctuation spectra depending on orientation perpendicular or parallel to the overall field. Thus, a threshold ratio for each component can be defined as a fraction of the total magnitude as in,
\begin{equation}
R_{j} \geq \frac{|B_{j}|^{2}}{|\vec{B}|^{2}}
\label{eq:Bthreshold}
\end{equation}
which reflects the relative amount that the total vector points in one of the three orthogonal directions. Then for every time, t, in a given time range and for each shot, the quantity $B_{j}(f,t)$ is summed for each t where
\begin{equation}
\frac{|B_{j}|^{2}}{|\vec{B}|^{2}} \geq R_{j}
\label{eq:Bcondition}
\end{equation}
for all frequencies, $f$. The value chosen for $R_{j}$ determines how strictly the total vector aligns with the component, $j$. By definition, then, the summed power for $B_{j}$ is considered the parallel component and the sum of the remaining two directions is the perpendicular component. If there is any anisotropy in the signal, a difference in energy content of the spectra should become apparent as the threshold value is increased.

\begin{figure}[!htbp]
\centerline{
\includegraphics[width=8.5cm]{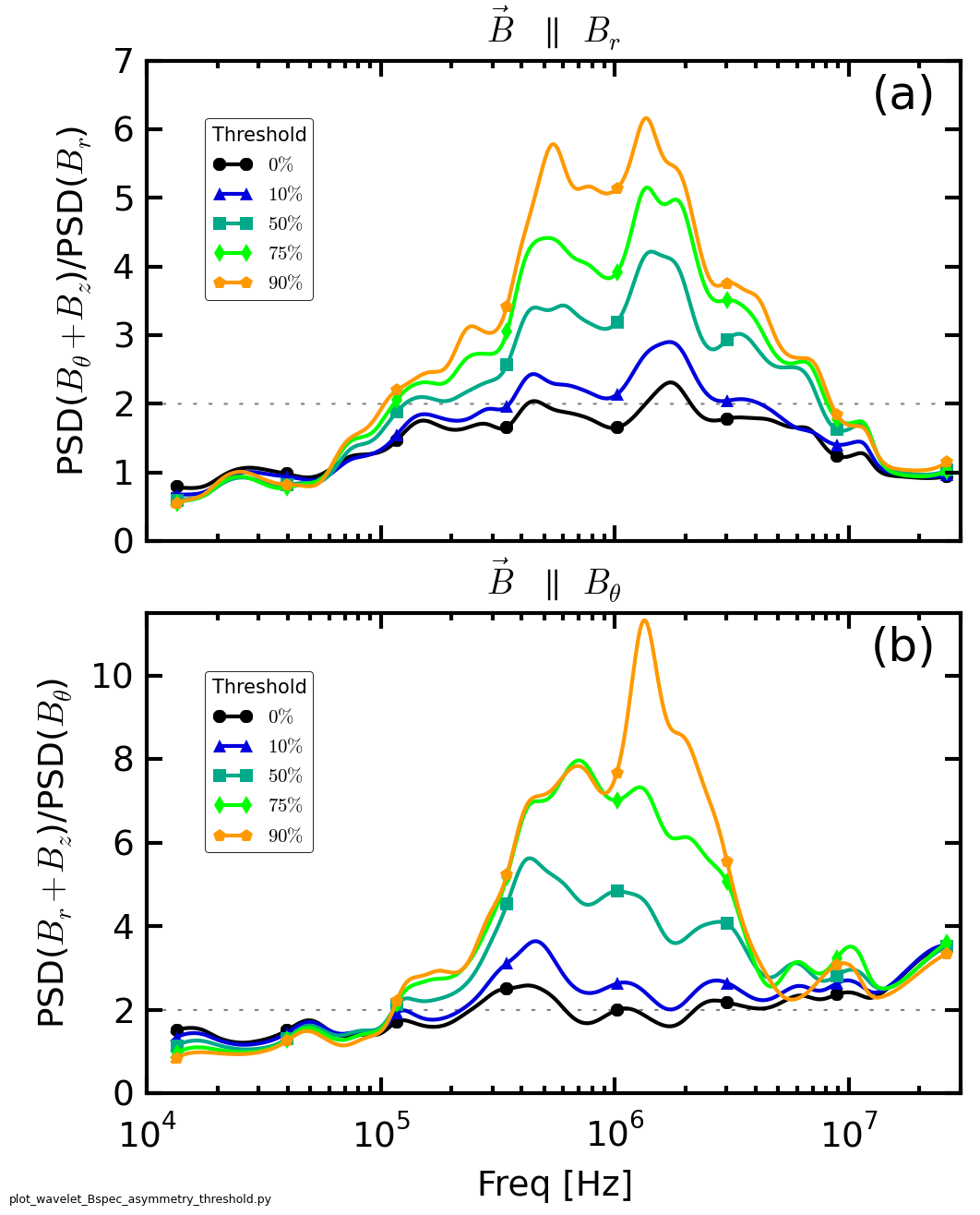}}
\caption{\label{fig:thresholdmethod} (a)Ratio of the sum of magnetic power spectra of $B_{\theta}$ and $B_{z}$ over the spectra of $B_{r}$ as a function of frequency and summed over the number of instances in time when the threshold ratio, defined in Equation~\ref{eq:Bthreshold} for $j=r$, is surpassed. Each curve indicates the level of anisotropy for different threshold values. The threshold ratio indicates the relative amount the full magnetic field vector, $\vec{B}$ points in the direction of $B_{r}$. (b) Similar to (a), but for $j=\theta$.}
\end{figure}

Indeed, an effect like this is observed. Figure~\ref{fig:thresholdmethod} shows the ratio of total perpendicular fluctuation power to parallel fluctuation power for $j = r$(Figure~\ref{fig:thresholdmethod}(a)) and $j = \theta$(Figure~\ref{fig:thresholdmethod}(b)). The threshold fraction is indicated by color. The dashed line at 2 represents isotropy---where the sum of 2 perpendicular components is about twice the power of the single parallel component. Clearly, for the lowest threshold value, the ratio remains close to the isotropy line for all frequencies as would be expected. As the threshold value is raised, the ratio from about 10kHz and higher begins to grow. This shows there is variance anisotropy in the plasma. If the plasma were isotropic, a difference between perpendicular and parallel spectra would not be seen. The anisotropy ratio reaches a maximum as the threshold nears 100\%. The drawback to this method, however, is that as the threshold is increased, the number of individual spectra summed is reduced which increases the error of each curve. This also probably explains why the threshold method indicates higher ratios then the more highly averaged data from the projection method discussed next.

\subsection{Projection Method}\label{sec:projection2}

An alternative method, and that which is used to compute the results presented in this manuscript, uses the $\vec{B}$ timeseries data to project spectral power into perpendicular and parallel portions at each timestep. This projection method uses all the available timesteps and shots rather than making a cut like the threshold method. It will be shown later that the two methods give quantitatively similar answers for the amount of variance anisotropy.

The projection method also uses the wavelet transform to compute $B_{j}(f,t)$. However, rather than use the $\vec{B}(t)$ as a threshold value, it is used as a reference vector to determine what fraction of the fluctuation power of $B_{r}(f,t)$, $B_{\theta}(f,t)$, and $B_{z}(f,t)$ is perpendicular or parallel to that vector. The parallel component of each $B_{j}(f,t)$ is found by computing the projection,
\begin{equation}
Proj_{u}v = \frac{\vec{v} \cdot \vec{u}}{||\vec{u}||}\vec{u}
\label{eq:projection}
\end{equation}
which shows that the magnitude of the component of $B_{j}$ parallel to $\vec{B}$ is
\begin{equation}
(B_{j}^{\parallel})^{2} = \frac{B^{2}_{j}}{|\vec{B}|^{2}}.
\label{eq:para_mag}
\end{equation}
Then, the magnitude of the component perpendicular is
\begin{equation}
(B_{j}^{\perp})^2 = 1 - \frac{B^{2}_{j}}{|\vec{B}|^{2}}.
\label{eq:perp_mag}
\end{equation}
Using these projection coefficients, each wavelet transform spectra, $B_{j}(f,t)$ for $j = r,\theta,z$, can be divided into $B^{\parallel}(f,t)$ and $B^{\perp}(f,t)$. For each timestep during each shot, the total parallel and perpendicular power is found by summing the respective portions from each orthogonal direction, as in
\begin{equation}
\begin{split}
B_{\parallel}(f) = \sum_{t=t_{0}}^{t_{1}} &(B_{r}^{\parallel})^{2}(B_{r}(f,t))+(B_{\theta}^{\parallel})^{2}(B_{\theta}(f,t))\\
																					&+(B_{z}^{\parallel})^{2}(B_{z}(f,t))
\label{eq:Bparallel}
\end{split}
\end{equation}
\begin{equation}
\begin{split}
B_{\perp}(f) = \sum_{t=t_{0}}^{t_{1}} &(B_{r}^{\perp})^{2}(B_{r}(f,t))+(B_{\theta}^{\perp})^{2}(B_{\theta}(f,t))\\
																					&+(B_{z}^{\perp})^{2}(B_{z}(f,t))
\label{eq:Bperp}
\end{split}
\end{equation}
for the given time range $t_{0}\leq t \leq t_{n}$. These summed quantities, $B^{\parallel}(f,t)$ and $B^{\perp}(f,t)$ are what are used in Section~\ref{sec:variance}.

\begin{figure}[!htbp]
\centerline{
\includegraphics[width=8.5cm]{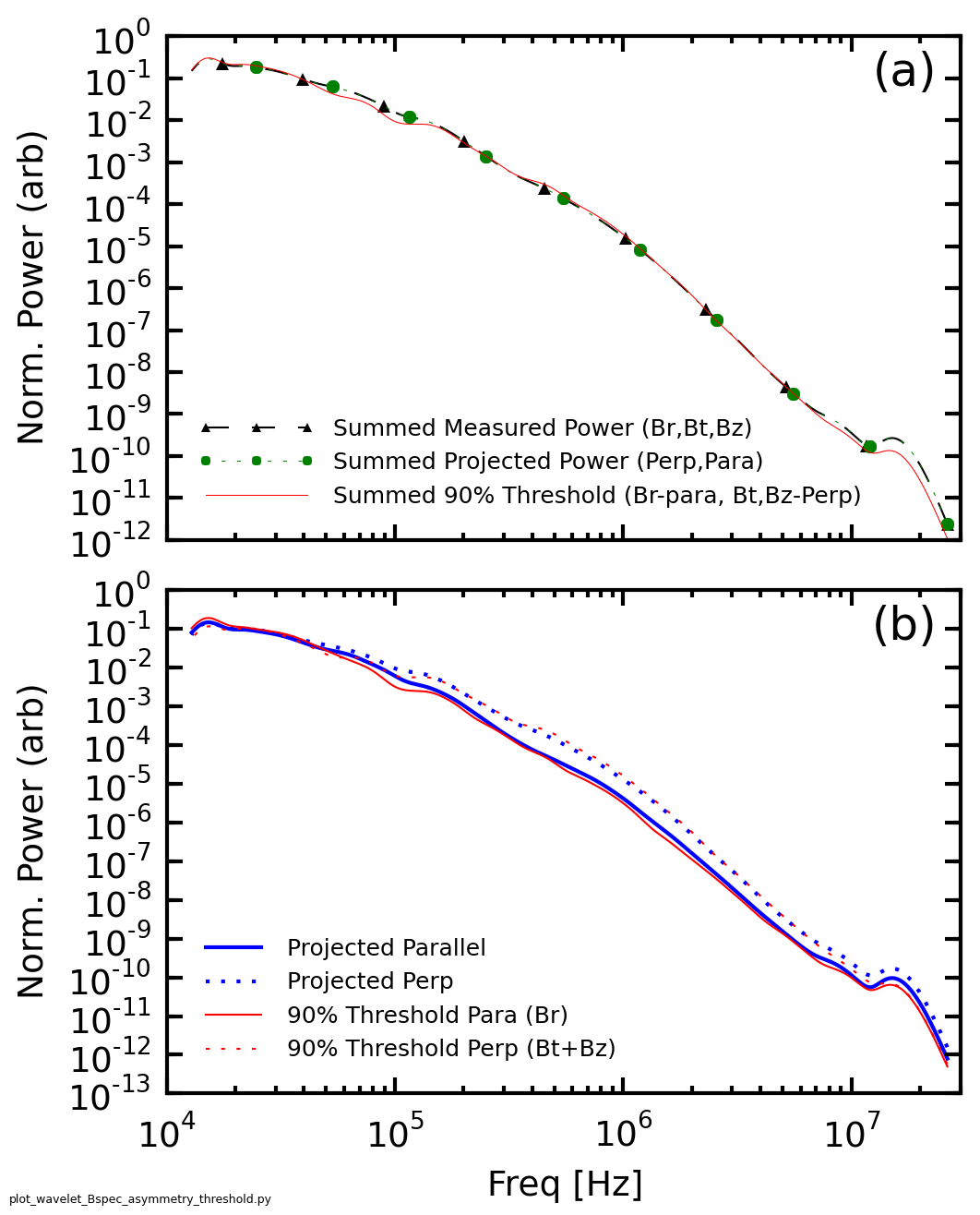}}
\caption{\label{fig:powercomparison} (a) Comparison of the total summed magnetic power showing the exact correspondance between ($B_{r},B_{\theta},B{z}$) and ($B_{\parallel},B_{\perp}$). Red curve shows the total power for the 90$\%$ threshold level which while not exact, is reasonably close to the other two curves. (b) Comparison of anisotropy in magnetic spectra using the projected method and the threshold method.}
\end{figure}

As a check, the total fluctuation power spectrum is computed in three different ways and plotted in Figure~\ref{fig:powercomparison}(a) for a time range of 40 to 60$\mu s$. The total power is found by (1) summing $B_{r}(f)$, $B_{\theta}(f)$, and $B_{z}$ directly, (2) summing $B^{\parallel}(f)$ and $B^{\perp}(f)$, (3) and summing the 90\% threshold spectra of $B_{r}$, $B_{\theta}$, and $B_{z}$ from the threshold method. The curves are averaged over the total number of timesteps used in their construction so they can be directly compared amongst one another. The first two ways match exactly, showing that the total power is being properly portioned. The 90\% threshold calculation does not match exactly, though is close. Figure~\ref{fig:powercomparison}(b) shows a comparison of the variance anistropy of the frequency power spectra as computed by the threshold and the projection methods. Again, the curves are normalized to total number of timesteps used in construction. The quantitative comparison shows that the projection method works well to compute the level of anisotropy as it compares well to the more direct, robust threshold method.

\section{Wavelet vs. FFT}\label{sec:WaveFFT}

\begin{figure}[!htbp]
\centerline{
\includegraphics[width=8.5cm]{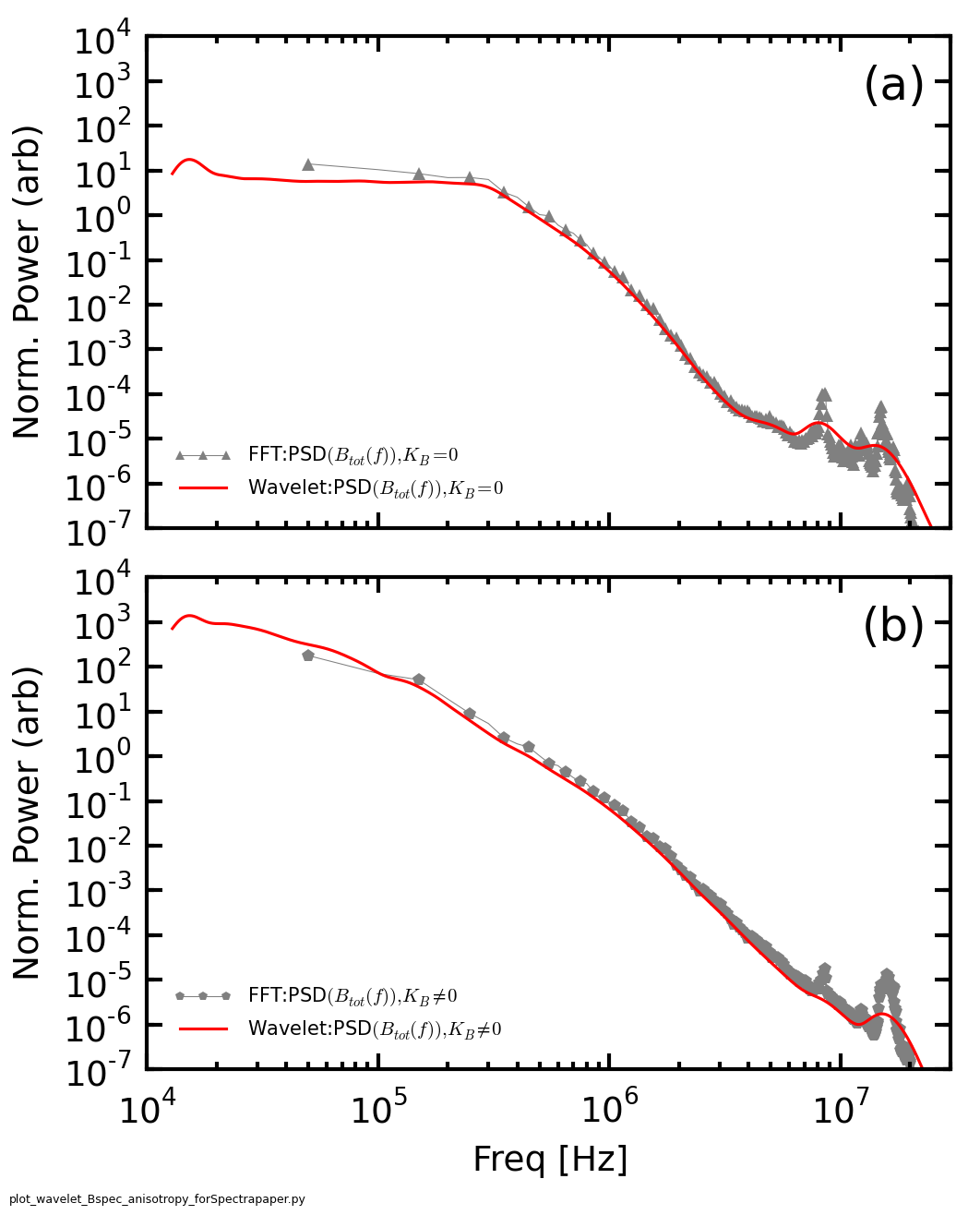}}
\caption{\label{fig:WavevsFFT} Comparison of magnetic power spectra constructed using a wavelet transform(red) and an FFT(gray,triangles) for zero helicity(a) and non-zero helicity(b) states.}
\end{figure}

Figure~\ref{fig:WavevsFFT} demonstrates the correspondance between magnetic fluctuation spectra using the wavelet analysis discussed in Section~\ref{sec:analysis} and a traditional Fast Fourier Transform. The red curves in both Fig~\ref{fig:WavevsFFT}(a) and (b) are the wavelet generated spectra using the full 120$\mu s$ of each shot for the zero and non-zero helicity states respectively. The gray curves show a Fourier transform generated spectra for only the 40-60$\mu s$ range of each shot. Clearly, the overall shape between the two sets of curves is nearly identical. Deviations occur at low frequencies where the wavelet transform can sample slightly lower frequencies since the analysis uses the entire time range rather than a time subset. The nature of the wavelet transform also allows for higher resolution binning especially at these lower frequencies. The wavelet transform, especially with the particular mother wavelet used---Morlet---tends to cause some smoothing in frequency space compared with the FFT. This is clearly seen at higher frequencies as a modes around 10MHz are more clearly observed in the FFT curve than the wavelet. However, for this particular dataset, these modes are not pertinent as they are caused by characteristics of the gun system and not by turbulence physics.

\section{Local vs. Global Magnetic Field}\label{sec:LocGlobe}

\begin{figure}[!htbp]
\centerline{
\includegraphics[width=8.5cm]{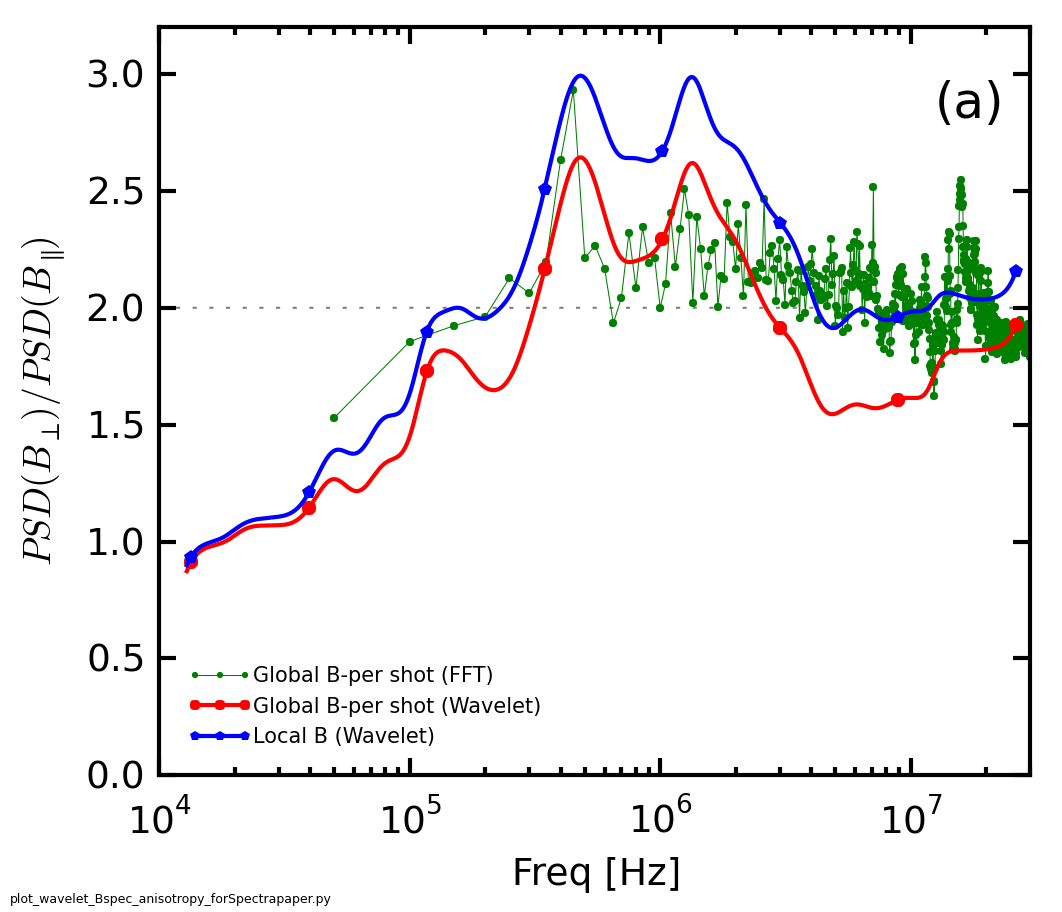}}
\caption{\label{fig:globalcomparison} Comparison of the anisotropy ratio for magnetic power spectra constructed when using either a local(blue,pentagons) or global(red,circles) definition of the mean field. The thin green curve shows the anisotropy with a global mean field using an FFT transform rather than wavelet.}
\end{figure}

The use of a temporally local versus global magnetic field in anisotropy analysis in the solar wind has been often debated~\cite{podesta09,matthaeus12}. In this paper, a local magnetic field reference vector has been used, but a relative global field can also be used to establish anisotropy. Since the experimental data is extracted on a shot-by-shot basis, the global field in this case is the mean field for the time duration of each shot. Figure~\ref{fig:globalcomparison} shows the anisotropy ratio for local field (reference vector at each timestep) and global field (mean field for each shot). While the local field yields a ratio that is slightly higher than for the global, the trend as a function of frequency is clearly similar. This comparison can be extended to the FFT analysis which typically would not be useful for a variance analysis technique because it does not have the time resolution that the wavelet transform does. However, since the magnetic field does not change very quickly in these plasmas, a mean field can be chosen for each shot and then used to project the full FFT spectra generated for each shot. The green curve shows this ratio, which while not as distinct as the wavelet generated ratio, nevertheless shows a similar trend especially in the kHz to MHz range. These results all suggest that though the use of a global versus local field may modify the exact numerical relationship, an anisotropy trend can be observed in either case if it is present in the plasma.
% ----------------------------------------------------------------
\section*{Acknowledgements}
  We gratefully acknowledge many useful discussions with William Matthaeus, Greg Howes, Kris Klein, Robert Wicks, Jason TenBarge and Adrian Wan. This work has been funded by DOE OFES and NSF CMSO.  The simulations were performed using the advanced computing resources (Cray XC30 Edison system) at the National Energy Research Scientific Computing Center.
% ----------------------------------------------------------------

\end{document}